\documentclass{aa}
\usepackage{graphicx,natbib}
\usepackage{txfonts}

\begin{document}
\title{The star formation history of luminous infrared galaxies}

\subtitle{}


\author{D. ~Marcillac\inst{1,8},
D. Elbaz\inst{1},
S. Charlot\inst{2,3},
Y. C. Liang\inst{4,5},
F. Hammer\inst{5},
H. Flores\inst{5},
C. Cesarsky\inst{6},
A. Pasquali\inst{7}}

\offprints{D. Marcillac, email: marcilla@as.arizona.edu}

\institute{DSM/DAPNIA/Service d'Astrophysique, CEA/SACLAY, 91191 Gif-sur-Yvette Cedex, France\\
\email{marcilla,elbaz@cea.fr}
\and
Max-Planck-Institut f\"ur Astrophysik, Karl-Schwarzschild-Strasse 1, Garching D-85748, Germany
\and
Institut d'Astrophysique de Paris, CNRS, 98 bis Boulevard Arago, Paris 75014, France\\
\email{charlot@iap.fr}
\and
National Astronomical Observatories, Chinese Academy of Sciences, No.20A Datun Road, Chaoyang District, Beijing 100012, P. R. China\\
\email{ycliang@bao.ac.cn}
\and
GEPI, Observatoire de Paris, Section de Meudon, 92195 Meudon Cedex, France\\
\email{hector.flores,francois.hammer@obspm.fr}
\and
ESO, Karl-Schwarzschild Strase 2, D85748 Garching bei Munchen, Germany\\
\email{ccesarsk@eso.org}
\and
Max-Planck-Institut fuer Astronomie, Koenigstuhl 17, D-69117 Heidelberg, Germany\\
\email{pasquali@phys.ethz.ch}
\and Current address: Steward Observatory, University of Arizona, 933, N. Cherry Avenue, Tucson, AZ 85721, USA\\
\email{dmarci@as.arizona.edu}
}

   \date{Received ...; accepted ...}

 
  \abstract
   {}
   {We constrain the past star
formation histories of a sample of 25 distant ($\bar{z}\sim$ 0.7) luminous
infrared galaxies (LIRGs) detected with the mid infrared cameras ISOCAM
and MIPS onboard the ISO and Spitzer satellites.}
   {We use high resolution VLT-FORS2 spectroscopy in addition to a
comprehensive library of 200,000 model optical spectra to derive Bayesian
likelihood estimates of the star formation histories of these galaxies based on the analysis of Balmer absorption lines and the 4000\,\AA $\,$break.}
   { The locus of distant LIRGs in the diagram defined by H$\delta_A$  and D4000 is roughly comparable  to that
of local LIRGs observed with IRAS, suggesting that no trend for an  evolution is detected between the local and distant LIRGs. 
We obtain similar results when
using either the H8 or the
H$\delta_A$ Balmer absorption-line indices in combination with D4000.

We compute a birthrate parameter (b=SFR/$<$SFR$>$) of 4$\pm$1,
confirming that the distant LIRGs are currently experiencing a
major phase of star formation.
The most likely
duration of the bursts is 0.10$^{+0.16}_{-0.06}$ Gyr, during which 
the LIRGs
produce $\sim$5-10\,\% of their current stellar mass. 
No evidence is found for successive starbursts on the scale of a
few times $10^7$\,yr, such as those predicted by some numerical
simulations of major mergers. 
However, the large number density of those galaxies suggest that  they could experience between two and four LIRG phases until the
present epoch.
This scenario is not consistent with the formation of the 
z$\sim$0.7 LIRGs through the continuous star formation characterizing
isolated spiral galaxies as has been argued independently based on
their morphology. Instead, minor mergers, tidal interactions or gas
accretion remain plausible triggering mechanisms for more
than half of the distant LIRGs which do not harbor the morphology of
major mergers.}
{}

\titlerunning{Star formation history of distant LIRGs}
\authorrunning{Marcillac et al.}
\maketitle
\section{Introduction}
Luminous infrared (IR) galaxies, i.e. galaxies radiating more than $\sim$ 90\,\% of their light above 5\,$\mu$m, have been suggested to provide important constraints on galaxy formation and evolution. They are considered to be the main cause of the cosmic infrared background (CIRB) and major contributors to the evolution with redshift of the cosmic star formation rate (CSFR) of galaxies (Elbaz et al. 2002, Chary \& Elbaz 2001, see also the review by Lagache {\it{et al.}} 2005). The role of luminous IR galaxies (LIRGs for $10^{11}$L$_{\sun}\leq L_{\rm IR}=~L[8-1000\,\mu{\rm m}]< 10^{12}$L$_{\sun}$) and ultra-luminous IR galaxies (ULIRGs for $L_{\rm IR}\geq 10^{12}~$L$_{\sun}$) in the local universe could be neglected since they only produce about 6\,\% of the integrated IR emission of local galaxies, hence 2\% of their bolometric luminosity (Soifer $\&$ Neugebauer 1991). But they rapidly evolve with redshift and dominate the CSFR above $z\sim$ 0.6 (Le Floc'h et al. 2005). IRAS extragalactic surveys already suggested that their number density evolved rapidly with redshift ($\sim~(1+z)^{7.6 \pm 3.2 }$ up to $z\sim0.2$; Kim $\&$ Sanders, 1998). This rapid evolution was later on confirmed up to $z\sim$ 1 with ISOCAM and ISOPHOT onboard ISO at 15 and 170\,$\mu$m (see Elbaz et al. 2005 and references therein), above $z\sim$ 2 using SCUBA at 850\,$\mu$m (Smail et al. 2001, Blain et al. 1999) and in the intermediate redshift range with MIPS onboard Spitzer at 24\,$\mu$m (Chary et al. 2004, Papovitch et al. 2004, Le Floc'h et al. 2005). However, little is known about the characteristics of the starbursts themselves, e.g. the amount of stars born during the burst, the burst duration, and even less about the physical processes responsible for the intense activity of these galaxies. One of the major reasons for this lack of information comes from their nature itself, i.e. the strong dust obscuration of their optical light. 

In a previous paper (Liang et al. 2004, hereafter Paper I), we presented an analysis of the emission line properties of the galaxies. The star formation rates (SFR) derived from the Balmer emission lines ($H\alpha$ and/or $H\beta$, plus $H\gamma$ to derive dust attenuations) were corrected for dust attenuation and found to be consistent with the ones derived from the mid IR (MIR) using the technique described in Chary \& Elbaz (2001). This study showed that LIRGs in general are not completely obscured by dust and that the use of high-resolution optical spectroscopy ($\Delta \lambda$/$\lambda$ = 2000, in the rest-frame of the objects) could be used to derive intrinsic luminosities, hence SFR, in rough agreement with the IR-derived SFR, by minimizing the contamination by sky emission lines and allowing to better correct for underlying photospheric absorption lines. However, the consistent derivation of the signal-to-noise ratio on the intrinsic luminosities lead to large uncertainties on the measured visual attenuation. Moreover, there is evidence for some completely obscured star formation as found in the most luminous objects studied in Flores et al. (2004) or in Cardiel et al. (2003). The limited statistics of those studies clearly calls for an extention of the sample of distant LIRGs, with good S/N on the optical continuum and high spectral resolution, to robustly determine which fraction of the star formation taking place in LIRGs and ULIRGs is completely obscured by dust. However Hopkins {\it{et al.}} (2003) showed that SFR([OII]), SFR(1.4 GHz), and SFR(FIR) are in very good agreement for a larger sample of local infrared galaxies detected with IRAS and spectroscopically observed with the Sloan Digital Sky Survey (SDSS).

In the present paper, we wish to address the problem from another angle: stars less massive, hence with longer lifetimes, than those responsible for the emission lines standardly used to derive the optical SFR of galaxies can escape their parent giant molecular cloud (GMC) and their spectral signature might be used to derive key parameters concerning the starburst. The H$\alpha$ light used to derive a SFR is dominantly produced by the ionizing photons arising from stars more massive than $\sim$10 M$_{\odot}$, with lifetimes shorter than 3 Myr. Those stars never escape their parent GMC (average lifetime of 10 Myr) and the dense regions of very strong extinction, contrary to the A and F stars which are the main contributors to the Balmer absorption lines and the 4000\,\AA  break.

We use these signatures of the optical continuum to compare distant LIRGs to nearby IRAS galaxies or synthetic spectra generated with the Bruzual \& Charlot (2003) model. We find a signature of the starbursts in the D4000-H$\delta$ diagram and use it to derive the burst properties.

Paper I indirectly confirmed the strong role played by LIRGs in the CSFR history with the derivation of gas metallicities in distant LIRGs twice smaller than the one measured in present-day galaxies of equal absolute B band magnitude. This result suggested that these galaxies produced about half of their metals between $z\sim$ 1 and today. It was also suggested that such a large metal production, as well as the large contribution of LIRGs to the CSFR and CIRB, could not result from a single burst phase in the galaxies harboring LIRG phases and hence that those galaxies must have experienced a series of LIRG phases in their lifetime. This possibility is tested in the present paper.

Finally we note that, based on optical spectra, no evidence was found in Paper I for a dominant contribution from active galactic nuclei (AGNs) in the sample of 76 distant LIRGs for which a spectroscopic redshift was determined. This confirmed the previous result from Fadda et al. (2001) that AGNs were contributing to less than 20\,\% of the MIR light of distant LIRGs, as shown by their soft to hard X-ray radiation measured with the Newton and Chandra X-ray observatories.

In this paper, we study the stellar spectra of a sample of 25 LIRGs with a median redshift of $\bar{z}=$ 0.7 using high-resolution spectroscopy with the FORS2 instrument at the VLT ($\Delta \lambda$/$\lambda$ = 1200, equivalent to 2000 in the rest-frame of the galaxies). This sample is smaller than in Paper I because higher signal-to-noise ratios are required to study the continuum emission in comparison to the emission lines. All galaxies are detected at 15\,$\mu$m with ISOCAM and the 11 galaxies located in one of the three fields are also detected at 24\,$\mu$m with the MIPS camera onboard Spitzer (Papovich et al. 2004, Elbaz et al. 2005). We show that both indicators provide consistent estimates of the total IR luminosity of the galaxies, hence also the SFR (see also Elbaz et al. 2005, Marcillac et al. 2005 for a more detailed study). 

Sect.2 presents the sample selection and the wavelet decomposition technique that we used to analyze the spectra. Sect.3 describes the method used in this paper to study the SFR history of the galaxies, namely the Balmer absorption line H$\delta_A$ (4101 \AA) versus 4000\,\AA\,break position of the galaxies, as previously done by Kauffmann et al. (2003) for the Sloan Digital Sky Survey (SDSS). 
We extended the method used in Kauffmann et al. (2003) to the high order Balmer line H8 (3889 \AA) and H9 (3835 \AA)which are easier to detect in  distant galaxies due to the k-correction.
The comparison of local and distant LIRGs in this parameter space is discussed in Sect.4 while the model used to generate Monte Carlo realizations of 200,000 spectra with different star formation histories is presented in Sect.5. The results are presented in Sect.6 and discussed in the last Section.

Throughout this paper, we will assume H$_o$= 75 km s$^{-1}$
 Mpc$^{-1}$, $\Omega_{\rm matter}$= 0.3 and $\Omega_{\Lambda}= 0.7$.

\section{Sample selection and data reduction}
\begin{table*}
\centering
\begin{tabular}{lccccccccrrrrr}
\hline
\hline
Object& RA        & DEC          & z    & H8     & H$\delta_A$            & D4000        &F$_{\nu}^{15\,\mu{\rm m}}$ &F$_{\nu}^{24\,\mu{\rm m}}$ & log($\frac{L_{\rm IR}}{L_{\odot}}$) & SFR$_{\rm IR}$ \\
 Slit & (J2000)   & (J2000)      &      &        &             &              &$\mu$Jy&$\mu$Jy&  & $M_{\odot}$\,yr$^{-1}$ \\
(1)   &    (2)    &  (3)         &  (4) & (5)    & (6)          &   (7) & (8)             &  (9) & (10)  &(11)   \\
\hline
\hline
UDSR08&3 14 55.20 &$-55$ 20 31.0 &0.7291& 3.8$\pm$0.5 & 4.2 $\pm$    0.9               &1.11$\pm$0.02 & 236$\pm$75 & --       & 11.44$\pm$0.24 &  46.7  \\
UDSR09&3 14 56.10 &$-55$ 20 08.0 &0.3884& 1.4$\pm$0.4 & --                            &1.74$\pm$0.03 & 609$\pm$87 & --        & 11.20$\pm$0.08 &  27.2  \\
UDSR10&3 14 43.90 &$-55$ 21 35.0 &0.6798& 4.3$\pm$0.6 & 8.4  $\pm$    0.7               &1.25$\pm$0.03 & 495$\pm$91 & --      & 11.78$\pm$0.13 & 102.9  \\
UDSR14&3 14 43.30 &$-55$ 20 11.0 &0.8150& 4.7$\pm$0.4 & 3.9 $\pm$    0.4               &1.15$\pm$0.02 & 200$\pm$69 & --       & 11.48$\pm$0.27 &  52.0  \\
UDSR20&3 14 41.10 &$-55$ 18 40.0 &0.7660& 6.9$\pm$0.5 & 6.5 $\pm$    0.8               &1.19$\pm$0.02 & 214$\pm$72 & --       & 11.44$\pm$0.25 &  47.0  \\
UDSR23&3 14 32.10 &$-55$ 19 02.0 &0.7094& 4.9$\pm$0.4 & 4.7 $\pm$    0.3               &1.15$\pm$0.02 & 271$\pm$80 & --       & 11.48$\pm$0.22 &  51.8  \\
\hline
UDSF06&3 13 30.20 &$-55$ 04 04.0 &0.6928& 4.3$\pm$1.3 &      --                     &1.27$\pm$0.06 & 150$\pm$75 & 180$\pm$22   & 11.09$\pm$0.49 &  21.0  \\
UDSF07&3 13 17.30 &$-55$ 05 16.0 &0.7014& 3.6$\pm$0.9 & 4.2 $\pm$    0.7               &1.17$\pm$0.04 & 154$\pm$32 & 198$\pm$27& 11.12$\pm$0.15 &  22.6  \\
UDSF12&3 13 07.70 &$-55$ 05 26.0 &0.7388& 7.6$\pm$0.9 & --                           &1.29$\pm$0.06 & 331$\pm$58 & 420$\pm$30  & 11.64$\pm$0.11 &  74.7  \\
UDSF16& 3 13 08.0 &$-55$ 04 18.0 &0.4548& --          & 3.7$\pm$0.6                  &1.25$\pm$0.01 & 138$\pm$29&  205$\pm$26     & 10.55$\pm$0.11 &   6.0\\
UDSF17&3 13 08.60 &$-55$ 03 57.0 &0.8100& 6.2$\pm$0.3 & 6.7 $\pm$    0.4              &1.22$\pm$0.02 & 257$\pm$47 & 343$\pm$23 & 11.62$\pm$0.12 &  72.0  \\
UDSF18&3 13 16.5  &$-55$ 02 27.0 &0.4620& --          & 4.7 $\pm$ 0.8                 &1.17 $\pm$0.01& 170$\pm$44& --       & 10.68$\pm$0.10 &   8.2\\
UDSF19&3 13 09.80 &$-55$ 03 08.0 &0.5476& 4.8$\pm$1.0 & 3.9 $\pm$    1.2              &1.32$\pm$0.05 & 611$\pm$90 & 778$\pm$40& 11.62$\pm$0.09 &  71.4 \\
UDSF28&3 12 51.80 &$-55$ 02 57.0 &0.6612& 2.9$\pm$0.6 &---             &1.47$\pm$0.04 & 321$\pm$35 & 861$\pm$37                & 11.49$\pm$0.06 &  53.1  \\
UDSF29&3 12 50.20 &$-55$ 02 59.0 &0.6619& 4.1$\pm$1.3 & ---        &1.37$\pm$0.06 & 354$\pm$66 & 568$\pm$28                    & 11.54$\pm$0.11 &  59.6  \\
UDSF20&3 13 19.00 &$-55$ 01 42.0 &0.8424& 2.8$\pm$0.7 &     --                     &1.33$\pm$0.04 & 117$\pm$25 & 248$\pm$21    & 11.19$\pm$0.16 &  26.7  \\
UDSF31&3 12 44.00 &$-55$ 03 21.0 &0.6868& 4.5$\pm$1.1 &  5.6$\pm$    1.1              &1.44$\pm$0.06 & 193$\pm$38 & 174$\pm$28 & 11.24$\pm$0.14 &  29.9  \\
\hline
						       	 	  	     	 		  			             			   
CFRS02&3 02 52.01 &  +00 10 33.0 &0.6172& 6.9$\pm$1.0 &  5.9$\pm$     1.2              &1.30$\pm$0.10 & 582$\pm$114& --       & 11.75$\pm$0.12 &  95.8\\
CFRS06&3 02 49.08 &  +00 10 ~1.8 &0.6169& 3.8$\pm$0.4 &  5.0$\pm$     0.3              &1.21$\pm$0.02 & 370$\pm$95 & --       & 11.50$\pm$0.17 &  53.5\\
CFRS08&3 02 46.29 &  +00 13 53.6 &0.7154& 4.0$\pm$0.3 &  5.9$\pm$     0.9              &1.24$\pm$0.03 & 360$\pm$95 & --       & 11.65$\pm$0.17 &  75.7\\
CFRS10&3 02 44.57 &  +00 12 20.1 &0.5276& 5.6$\pm$0.6 &  5.9$\pm$     1.2             &1.03$\pm$0.04 & 220$\pm$95 & --        & 11.01$\pm$0.40 &  17.7\\
CFRS11&3 02 42.31 &  +00 12 9.5  &0.6895& --          &  2.5$\pm$     0.7             &1.15$\pm$0.04 & 425$\pm$110      &   --      & 11.70$\pm$0.16 &  85.4\\
CFRS14&3 02 40.44 &  +00 14 ~3.8 &0.4652& 4.6$\pm$0.5 &  6.5$\pm$     0.5             &1.10$\pm$0.02 & 172$\pm$117& --        & 10.69$\pm$0.58 &   8.4\\
CFRS16&3 02 38.80 &  +00 14 17.5 &0.8274& 3.3$\pm$1.0 &     --                     &1.30$\pm$0.05 & 444$\pm$100& --           & 11.99$\pm$0.15 & 166.7\\
CFRS29&3 02 29.41 &  +00 12 59.8 &0.8804& 7.3$\pm$1.4 &      --                    &1.13$\pm$0.05 & 370$\pm$103& --           & 11.98$\pm$0.20 & 163.6\\
						           
\hline
\hline
\end{tabular}
\caption{Description of the distant LIRGs sample. The technique used
to compute the equivalent widths of $H\delta$, $H$8, $H$9, $H$10 and
the 4000\,\AA~break ($D4000$) is described in
Sect.~\ref{SEC:defindice}. $L_{\rm IR}$ and
SFR$_{\rm IR}$ were derived using the Chary \& Elbaz (2001) technique.}
\label{TAB:echantillon}
\end{table*} 

\subsection{Sample selection}
The sample of distant LIRGs was selected from
three deep ISOCAM 15\,$\mu$m surveys and is
described in Paper I. 105 galaxies were selected on
the basis of their 15\,$\mu$m flux density in three different regions
of the southern hemisphere, hence avoiding strong contamination by
cosmic variance: the ISOCAM Ultra-Deep Surveys in the FIRBACK (UDSF)
and ROSAT (UDSR) fields and the CFRS 3$^h$ field. All three fields
were selected for their low cirrus contamination and high galactic
latitude.  The UDSF (9\arcmin$\times$9\arcmin) is located at the
center of the ``far infrared background'' survey at 175\,$\mu$m with
ISOPHOT onboard ISO ('FIRBACK'; Puget et al. 1999, Lagache \& Dole
2001).  The UDSR is centered at the position of a deep ROSAT survey
(Zamorani et al. 1999). The UDSR and UDSF are both close to the
position of the so-called 'Marano Field' originally selected for an
optical survey of quasars (Marano et al. 1988), but are separated by
21\arcmin with respect to each other.  The third field is one of the
Canada France Redshift Survey fields (CFRS 3$^h$) combining deep
infrared, optical and radio data as well as spectra from the MOS
multiobject spectrograph on the 3.6m CFHT.

ISOCAM sources were selected to span the whole flux density
range of the three surveys whose 80\,$\%$ completeness limits are
150\,$\mu$Jy for the UDSF and UDSR and 250\,$\mu$Jy for the CFRS
3$^h$, while the detection limits are 50\,$\mu$Jy (UDSF, UDSR) and
170\,$\mu$Jy (faintest sources in the regime below completeness). 
No optical selection was applied so that the magnitudes of the sources 
range from $R_{\rm AB}$= 19.4 to 23.7. At about the same epoch (November 2003) than 
the VLT-FORS2 observations, a 24\,$\mu$m survey was being performed with the MIPS camera  
onboard Spitzer during the MIPS commissioning
phase (IOC/SV). This survey covers the whole UDSF field and all eleven ISOCAM-15\,$\mu$m sources were detected at 24\,$\mu$m, i.e. 45\,\% of the sample.
The scan map AOT was used, with an
half-array overlap to cover about 1300 Sq. Arcmin with high redundancy
(20) and to get an integration time per sky pixel of about 230s
(Papovich et al. 2004). The data were reduced using the {\it Spitzer}
Science Center Pipeline and the BCD products (Basic Calibrated Data, Pipeline
version S10.0.3) were downloaded from the {\it Spitzer}
archive\footnote{http://archive.spitzer.caltech.edu, PID: 718}. 
 PSF-fitting photometry was performed using DAOPHOT
(Stetson 1987) with IRAF\footnote{IRAF is distributed by the National
Optical Astronomy Observatories, which are operated by the Association
of Universities for Research in Astronomy, Inc., under the cooperative
agreement with the National Science Foundation}.

From a total of 105 galaxies observed with FORS2 at the VLT (Paper I), we detected 3 stars, 13 galaxies were too faint for a redshift determination and within the remaining list, 13 other galaxies were not detected with ISOCAM. The resulting list of 76 galaxies with a MIR detection is divided into 34 "normal" galaxies, 36 LIRGs and 6 ULIRGs, where "normal" galaxies consist in all galaxies except IR luminous galaxies (i.e LIRGs and ULIRGs). The median SFRs associated to each total IR luminosity bin are 4, 54 and 196\,M$_{\odot}$ yr$^{-1}$ respectively. Because good quality spectra are required to study the stellar absorption lines (S/N$>$ 3 on the continuum per resolution element), the final sample that we study in the present paper consists of 25 LIRGs with IR luminosities ranging between 10$^{11}$ and 10$^{12}$\,L$_{\odot}$. The median redshift of the final sample is $\bar z$=0.7.
The measured properties of the galaxies are summarized in Table~\ref{TAB:echantillon}.

The total (8-1000\,$\mu$m) IR luminosities, $L_{\rm IR}$, were derived using
the library of template SEDs built by Chary \& Elbaz (2001), as
in Elbaz et al. (2005). We also computed $L_{\rm IR}$ from the Dale \& Helou (2001) library following the technique described in Marcillac et al. (2005) and found a median value for $L_{\rm IR}$ 10\,\% lower than with the previous library and with an rms of 17\,\%, hence both techniques provide consistent luminosities.
We then compared the $L_{\rm IR}$ derived with 15\,$\mu$m and/or 24\,$\mu$m flux densities for the 11 galaxies detected with both ISOCAM and MIPS. Both libraries of template SEDs provide consistent determinations of $L_{\rm IR}$ using both measurements (with an rms of 30\,\%, see also Elbaz et al. 2005). The median $L_{\rm IR}$ derived from MIPS is 10\,\% lower than the one derived from ISOCAM using both libraries which suggests a possible variation of the MIR spectra of LIRGs as a function of redshift (see Marcillac et al. 2005).

\subsection{Observations and data reduction}
\begin{figure*}
\resizebox{\hsize}{!}{\includegraphics{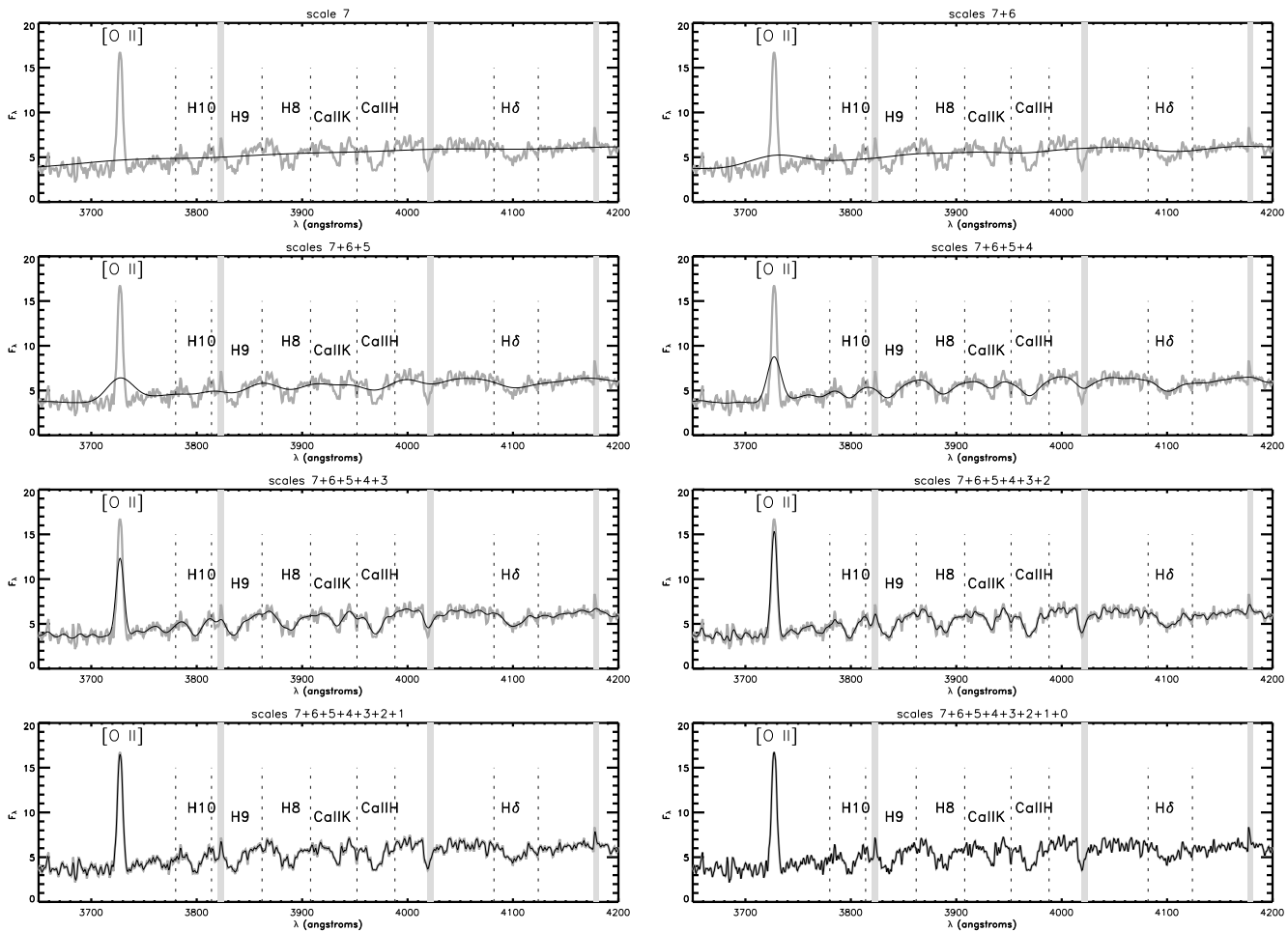}}
\caption{Wavelet decomposition of the VLT-FORS2 spectrum of a distant
LIRG (UDSR23) located at a redshift of z=0.7094) and forming stars at
a rate of SFR$_{\rm IR}$= 51.8 M$_{\odot}$ yr$^{-1}$. The raw spectrum
(thick grey line) is compared to its wavelet decomposition (thin dark
line), starting from scale 7, corresponding to the spectrum smoothed
with a B-spline (equivalent to a gaussian) at the lowest resolution of
89.6\,\AA, then adding up the wavelet scales by increasing the
spectral resolution from 44.8\,\AA (scale 6) to 0.7\,\AA (scale 1), by
dividing it by a factor 2 for each scale. The narrow vertical grey
zones were masked in the analysis because of the presence of sky
emission lines. The vertical dotted lines indicate the minimum and
maximum wavelength for each absorption line. In this zoom, the
absorption lines clearly appear as much shallower than the emission
line in OII.}
\label{FIG:decomp}
\end{figure*}

\begin{figure}
\resizebox{\hsize}{!}{\includegraphics{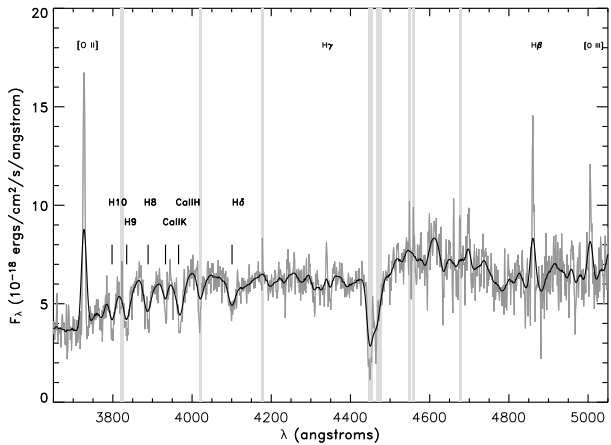}}
\caption{Comparison of the VLT-FORS2 raw spectrum (grey line) of the
distant LIRG UDSR23 (z=0.7094, 51.8 M$_{\odot}$ yr$^{-1}$) to its
wavelet decomposition resulting from the combination of the four
wavelet scales 4-7, i.e. from 11.2 to 89.6\,\AA (black line). While
the absorption lines are clearly distinguishable, the emission lines
such as [OIII],[OII],H$\beta$,H$\gamma$ are partly diluted by this
technique.}
\label{FIG:typSPEC1}
\end{figure}
The observations were performed during three nights with FORS2 on the
ESO-VLT with the combination of the grisms R600 and I600 (3 hours per grism) to cover the
wavelength range 5000 to 9200\,\AA \, at a resolution of 5\,\AA \,(R=1200).
At the median redshift of the objects of $\bar{z}\sim$0.7, the resolution is
equivalent to 3\,\AA \,(R=2000). 

Because absorption
lines are wider than emission lines, it is possible to increase the
signal-to-noise ratio on the absorption lines by working at a lower frequency
than the raw spectrum.  The physical origin of the width of the absorption lines results from the complex combination of the internal dynamics of the stars and the global dynamics of the galaxy. In order to optimize the line extraction, we therefore decomposed the raw spectra into eight different wavelet scales (Table~\ref{TAB:wave})
using the undecimated (keeping an identical sampling in
each wavelet scale) wavelet transform (\`a trous algorithm; Starck \&
Murtagh 1994, Starck, Pantin \&
Murtagh 2002). We optimized the S/N ratio on the absorption line
features by selecting the best combination of wavelet scales. In the
wavelet space, the first scale (highest frequency), that we index as
scale 0, corresponds to features of size 0.7\,\AA, while features at
scale $j$ have a size of 0.7$\times$2$^{\rm j}$\,\AA. Note that the
lowest frequency scale is equivalent to the raw spectrum smoothed by a
B-spline (equivalent to a gaussian) of width 89.5\,\AA, i.e. the
baseline, while at higher frequencies, each wavelet scale 'i' is equal
to the difference between the raw spectrum smoothed at the scales 'i'
and 'i-1'. Hence the sum of all wavelet scales plus the baseline (here
at scale 7) is exactly equal to the initial raw spectrum.

\begin{table}
\begin{center}
\begin{tabular}{cccc}
\hline
\hline
wavelet scale&resolution (\AA)\\
\hline
\hline
0&0.7\\
1&1.4\\
2&2.8\\
3&5.6\\
4&11.2\\
5&22.4\\
6&44.8\\
7&89.6\\
\hline
\hline
\end{tabular}
\end{center}
\caption{Definition of the wavelet scales used in the wavelet
decomposition of the VLT-FORS2 spectra. The spectral resolution is
equal to the highest resolution of 0.7$\times$2$^{\rm wavelet scale}$\,\AA. See description in the text for the construction of
the different wavelet transforms at each scale.}
\label{TAB:wave}
\end{table}

In order to determine the number of scales to take into account in the
decomposition, we started from the lowest resolution in the wavelet
space, equivalent to a spectral resolution of 89.5\,\AA. We then kept doubling the frequency level as long as the S/N ratio was increased.
  This happened at the wavelet scale '4' (equivalent to
11.2\,\AA), hence we used the co-addition of the 4,5,6 and 7 wavelet
scales (from 11.2 to 89.6\,\AA, increasing the frequency by a factor 2
for each scale) in order to reconstruct spectra devoid of high
resolution noise. The steps of the decomposition are shown in
the Fig~\ref{FIG:decomp} for a LIRG, UDSR23 ($z$= 0.7094), located at the median redshift of the sample. The spectra resulting from the combination
of these four scales present the advantage of seeing the same spectral
resolution as the one used by Bica $\&$ Alloin (1986). The final result is compared to the raw spectrum in the
Fig.~\ref{FIG:typSPEC1}. 

Note that the wavelet decomposition can dilute emission lines and that we have checked that absorption lines were not affected by a similar effect. Being wider, they are naturally less affected by this technique. However, we quantified this effect using some stellar spectra extracted from STELIB (Le Borgne et al 2003) which resolution is about the same as the rest-frame one for the distant LIRGs. 
We added a white noise to the stellar spectra to reach a signal to noise ratio of S/N $\sim$ 3 on the continuum, interpolated the spectra to reach a resolution of 0.7 $\AA$ and applied to them the same wavelet decomposition as for the distant LIRGs. The equivalent widths of the Balmer absorption lines determined before and after the wavelet analysis differ by less than 4\%.  We included this weak difference in the equivalent width uncertainties.
 
\subsection{Measurement of the Balmer absorption line indices and 4000\,\AA break}
\label{SEC:defindice}
In the following, we will compare the equivalent widths measured for $H\delta$ (4101\,\AA) and for the high-order Balmer absorption line H8 (3889 $\AA$). We also considered using the H9 (3835 $\AA$) line that we therefore also define in this section, but it appeared to be a less reliable tracer of the past star formation activity of galaxies, as we discuss in in the next section.

We used the definition of the $H\delta$ pseudo--equivalent width indice as
defined in the Lick system (Worthey $\&$ Ottaviani 1997)(see table~\ref{TAB:indices}).
Since no Lick indices have yet been defined for the high order Balmer absorption lines, we adapted the windows defined by Bica \& Alloin (1986) to the Lick index method for H8 and H9 : for each line, the index continuum, the blue and red bandpasses for each pseudo--continuum are summarized in table~\ref{TAB:indices}.
The two latter lines present the advantage of being located at a lower
wavelength, to be accessible to higher redshifts in the observed
optical range and to be less affected by the overlying nebular emission
lines at the same wavelengths. 

\begin{table}
\begin{center}
\begin{tabular}{cccc}
\hline
Name& Blue and red bandpasses (\AA)& Index continua(\AA)\\
\hline
\hline
$H\delta_A$    &4041.60-4079.75      4128.50-4161.00     &4083.50-4122.25  \\
H9           &3810-3820      3855-3865     &3825-3845 \\
H8           &3855-3865      3905-3915     &3870-3900 \\
\hline
\end{tabular}
\end{center}
\caption{Definition of the pseudo--equivalent width indices for the
$H\delta$ ($H\delta_A$ in Worthey $\&$ Ottavianni 1997) and for the
two high order Balmer absorption lines. For the two last lines,
we followed the same principle as for Lick indices while using
windows defined Bica $\&$ Alloin (1986).
}
\label{TAB:indices}
\end{table} 
The $H\delta$ absorption line is surrounded by iron absorption lines which affect
both its red and blue pseudo continua and are responsible
for the negative EW measured for this line after a few Gyears, as discussed in the next section.
In order to avoid such pollution from neighbouring lines, the red continuum of H8 was choosen to minimize the sensitivity to the CaII H (3933 \AA)and K (3966 \AA) lines while the 3855-3865 \AA $\,$ region is not known to be affected by metallic lines. As a result this index is weakly affected by metallicity.
H9 is the least polluted line because no metallic line is located
near it, but it is somewhat fainter than H8 and it is surrounded by the strong absorption of H8 and H10 which makes the two pseudo continua more difficult to define. We have primarily used H$\delta_A$ and $H8$, which are better
determined for most of our spectra but also checked that
we obtained consistent results based on the $H9$ and $H\delta$
lines when they were available. 
Note that we will discuss separately the results obtained with the Balmer absorption lines H$\delta_A$ (4102\,\AA) and H8 (3889\,\AA). 
The advantage of this choice is that it provides two independent estimates of the parameters that we are deriving and can be used as a test of the robustness of the Bruzual \& Charlot (2003) code that we are using. 

Note also that there is an ongoing debate about the
possible misinterpretation of the equivalent width of the H$\delta_A$ line because of metallicity ratios which could affect its neighboring regions, hence its associated pseudo-continua (see Thomas {\it{et al.}}, 2004, Korn {\it{et al.}}, 2005). On the one hand, H$\delta_A$ presents the advantage to allow the comparison with studies of local galaxies such as SDSS galaxies (Kauffmann et al. 2003) while H8 is usually not available for local galaxies because it lies in a bluer region of the spectrum. On the other hand, the H8 line is not known to be affected by neighboring metallic lines and as it is located in a bluer region of the spectrum, it is easier to measure for distant galaxies (less polluted by sky emission lines).

Before measuring these absorption
features, we have corrected them from the overlying nebular emission
line whenever possible as indicated in the paper. The nebular
emission lines are not detected directly from the spectra, because
they are too faint, but we computed their emission based on the
observed $H_{\beta}$ and $H_{\gamma}$ emission lines (and $H_{\alpha}$
for the low-$z$ galaxies) assuming a line ratio corresponding to a case B recombination for electron densities $\le$ $10^4$ $cm^{-3}$ and temperatures $T_e$ $\sim$ $10^4$ (Osterbrock, 1989). 
The Balmer
emission line ratio was also used to compute the attenuation of these
lines before subtracting them from the absorption lines measurements.
The computation and values for these attenuations can be found in
Paper I. 

Following the same strategy as Kauffmann et al. (2003), we used the method explained in Bruzual (1983) and the blue and red bandpass definitions introduced by Balogh et al. (1999). The latter are narrower than the ones
originally defined by Bruzual (1983) and present the advantage
of being less sensitive to reddening effects. 

H8 and D4000 values for each distant LIRG are summarized in 
Table~\ref{TAB:echantillon}. The scientific interpretation is
discussed in the following section.

\section{4000 $\AA$ break and Balmer absorption lines as tracers of the recent star formation history}
\label{SEC:tracing}
\begin{figure}
\resizebox{\hsize}{!}{\includegraphics{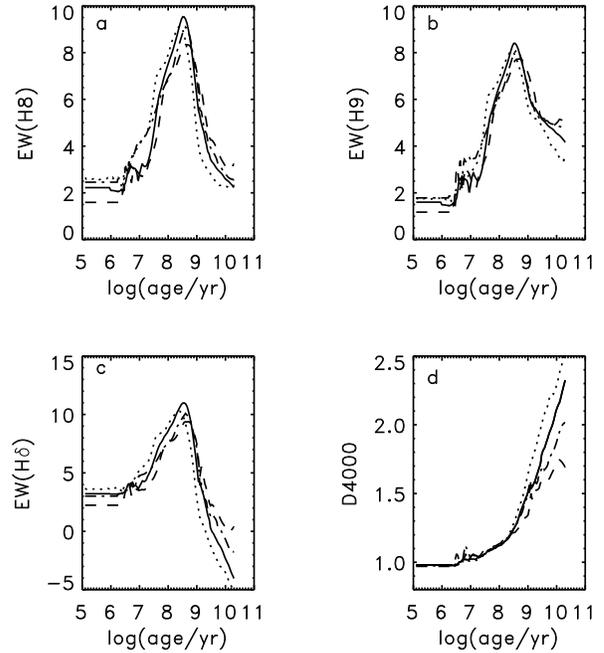}}
\caption{Time evolution of the H8, H9 and H$\delta_A$ indices and
4000\,\AA break (D4000) for a single stellar population synthesized
with the GALAXEV code (Bruzual \& Charlot 2003). Four different
metallicities were used in each plot demonstrating the marginal effect
of metallicity on the three indices and on D4000 before 1 Gyr: 
Z= 0.004 (20\,\% solar, dashed line), Z= 0.008 (40\,\% solar, dash-dotted line), Z= 0.02 (solar, solid line), Z= 0.05 (2.5 times solar, dotted line).  From top to bottom and left to right: a) H8 evolution,
b) H9 evolution, c) H$\delta_A$ evolution, d) D4000 evolution.}
\label{FIG:Tevol}
\end{figure}
In this section, we describe how Balmer absorption lines and the 4000 $\AA$ break trace the recent star formation history of galaxies. For this purpose, we have synthesized a single
stellar population (SSP) using the latest version
of the 'GALAXEV' code from Bruzual \& Charlot (2003). For the moment, we do not include dust attenuation to simplify the discussion. Note however that the wavelength range over which the equivalent widths and the 4000 $\AA$ break are measured is small which implies a marginal correction due to dust attenuation. Nonetheless, we will include dust attenuation in the Monte Carlo realizations that we will compare to the LIRGs and discuss its effects on our findings.

This version of the GALAXEV code includes the spectral library
STELIB (Le Borgne et al. 2003) whose spectral resolution is 3$\AA$ from
3200 to 9500 $\AA$, which is comparable to the present spectra in the
rest-frame of the galaxies. Fig.~\ref{FIG:Tevol} presents the
evolution with time of the high order Balmer absorption lines H8 and H9 as well as the H$\delta_A$ absorption line and the 4000
$\AA$ break. The four lines in each plot correspond to four different metallicities with the following metal mass fractions (total mass in elements heavier than hydrogen and helium over the mass in hydrogen): Z= 0.004 (20\,\% solar, dashed line), Z= 0.008 (40\,\% solar, dash-dotted line), Z= 0.02 (solar, solid line), Z= 0.05 (2.5 times solar, dotted line).

The D4000 is the sudden onset of stellar photospheric opacity shortward of 4000$\AA$.
It reflects the mean temperature of the stars responsible for the continuum: the metals located in the atmosphere of O and B stars are more ionized and produce a weaker opacity, hence a smaller 4000 $\AA$ break, than those in cooler stars (Bruzual 1983,
Poggianti \& Barbaro 1997, Gorgas et al 1999, Kauffmann et al 2003). Hence D4000 keeps increasing as a function of the aging of the stellar population (see Fig.~\ref{FIG:Tevol}d). D4000 is sensitive to metallicity as it is shown in tab.~\ref{FIG:Tevol}d where it varies by more than 20\,\% after a few billion years, or when it is larger than 1.6. As we will see in the next section, the distant LIRGs have D4000=1.2$\pm$0.07, and therefore for them metallicity effects are negligible. 
 Even if the slope of D4000 versus age is less flat than for young population, i.e. 7 Gyears, it is flat enough to provide uncertain stellar ages if used alone.

In order to trace back the recent star formation history of galaxies, it is therefore necessary to use
another tracer of stellar age such as the Balmer absorption lines, which exhibit a steep slope as a function of stellar ages in this range of ages (see Figs.~\ref{FIG:Tevol}a,b,c). Balmer absorption lines are dominantly produced by the atmosphere of A
to F stars. However O and B stars, which do not exhibit strong absorption lines, indirectly affect them by increasing the continuum level and therefore diluting them, which explains the flat values for the equivalent widths of H8, H9 and H$\delta_A$ in the first few millions years (lifetime of O and B stars). The rapid increase that follows is produced by the dominant role of A and F stars, which then disappears after $\sim$ 0.5 Gyr producing the rapid decline of the equivalent widths in Fig.~\ref{FIG:Tevol}a, b and c. Here again, it is worthwhile noticing the marginal role played by the metallicity on the evolution of the Balmer lines EW with time. The EW varies by less than 20\,\% a few billion years after the burst as a function of metallicity.

\section {Comparison of local and distant LIRGs}
\label{SEC:comp}
\begin{table}
\begin{center}
\begin{tabular}{ccc}
\hline
\hline
 indice    &distant LIRGs        &local LIRGs\\
\hline
\hline
D4000      & $1.2_{-0.07}^{+0.07}$&$1.25_{-0.08}^{+0.12}$\\
H$\delta_A$  & $5.0^{+1.5}_{-1.1}$ & $4.9_{-1.5}^{+1.07}$\\
\hline
\hline
\end{tabular}
\end{center}
\caption{Median and 68\,\% dispersion (around the median) of the D4000 and H$\delta_A$ indices for local and distant galaxies.}
\label{TAB:resultind}
\end{table}

\begin{figure*}
\resizebox{\hsize}{!}{\includegraphics{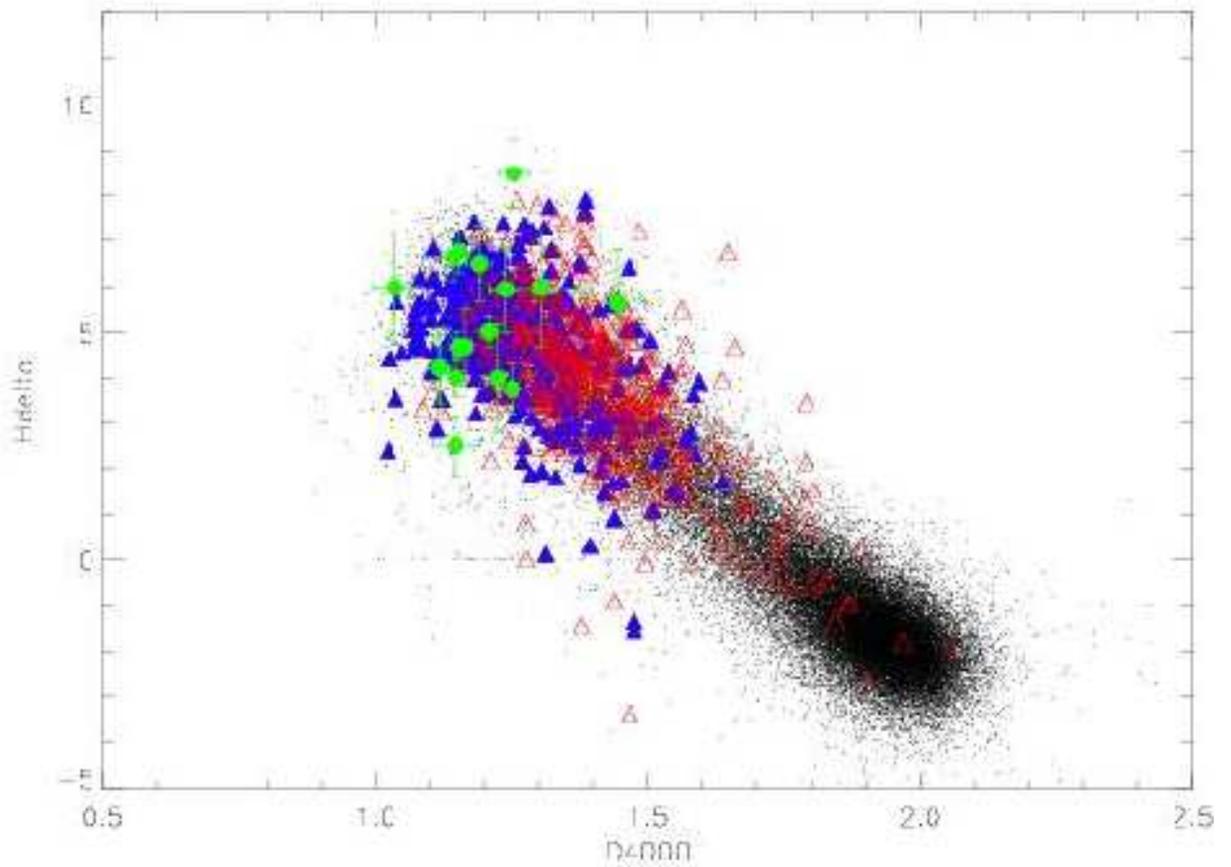}}
\caption{
H$\delta_A$ versus D4000 diagram for the SDSS galaxies (black points)
and distant LIRGs of the present sample (green filled circles with
error bars). The sub-population of SDSS galaxies detected with IRAS
are marked with triangles and separated into galaxies with optical
spectra with (empty red triangles) and without (filled blue triangles)
an AGN signature.
}

\label{FIG:irascam}
\end{figure*}

\begin{figure*}
\resizebox{\hsize}{!}{\includegraphics{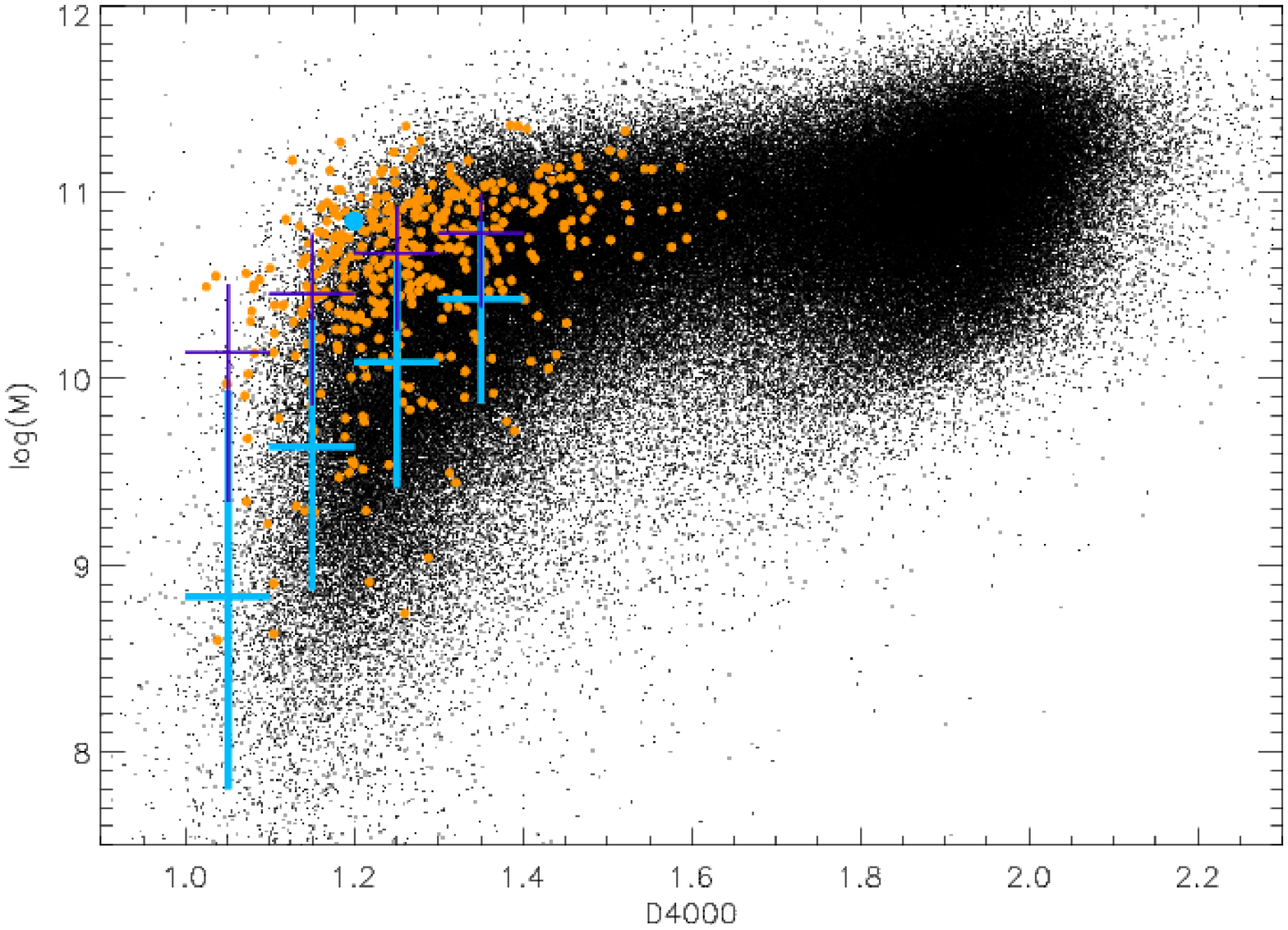}}
\caption{
Stellar mass as a function of D4000 for the SDSS galaxies (black
points) and distant LIRGs of the present sample (large filled blue
circle). The median stellar mass for the distant LIRGs was derived
from Franceschini et al. (2003) as discussed in the text. The
sub-population of SDSS galaxies detected with IRAS and without an AGN
signature in their optical spectra (star forming galaxies) are marked
with orange filled circles. The median and 1-$\sigma$ error bars for
the field SDSS galaxies are represented with thick light-blue crosses,
while the thin dark-blue crosses are for the sub-sample of star
forming SDSS-IRAS galaxies.}

\label{FIG:masses}
\end{figure*}
A sample of 401 local (z $\leq$0.25) LIRGs detected with IRAS, with optical spectra from the SDSS and emission lines typical of star forming galaxies, as opposed to AGNs, was identifed by Pasquali et al. (2005). The locus of the D4000 and EW (H$\delta_A$) measured for the distant LIRGs (green filled circles) is compared to that of local LIRGs (blue triangles) and of the field SDSS galaxies (black dots) in Fig.~\ref{FIG:irascam}. First, note the concentration of both LIRG populations at low D4000 and high H$\delta_A$, which suggests that local and distant LIRGs share a similar recent star formation history. We could not produce similar figures for high order Balmer absorption lines for this comparison since they are too blue to be accessible in the SDSS spectra. 
Their position in the D4000-H$\delta_A$ diagram indicates that the young stellar population that is producing the large IR luminosity is not completely obscured by dust in the optical because the median D4000 value of 1.2 for these galaxies correspond to stellar ages lower than 1 Gyr, which is much below the ages of these galaxies (see Table~\ref{TAB:resultind}).
This already suggests that even in these dusty galaxies the optical spectral signatures can be used as a tracer of the recent star formation history. The quasi absence of distant LIRGs above D4000= 1.25, where half of the local LIRGs lie, suggests that the distant LIRGs are younger than the local ones.

The relative proportion of young and old stellar populations can be studied in these galaxies by comparing the stellar masses, derived mostly from the old stellar population dominating in the near infrared range, with their D4000 values.
Massive galaxies are generally older as shown by Fig.~\ref{FIG:masses}, where the stellar mass of SDSS galaxies (black dots) increases with D4000. The position of the local LIRGs (orange filled dots) in Fig.~\ref{FIG:masses} suggests that they are massive galaxies which were located on the right side of the plot before the burst and which were shifted to the left during the burst, which decreased their D4000 value. This figure reinforces our interpretation that the low value of D4000 for the LIRGs is due to the addition of a young population on top of an older population. The comparison of field SDSS galaxies with IRAS LIRGs in four bins of D4000 from 1 to 1.4 (bin size 0.1, 68\,\% error bars) shows that local LIRGs exhibit systematically lower D4000 values for their stellar masses.

The incompleteness of the photometric data for our sample of galaxies
prevented the determination of their stellar masses.  However, another
sample of mid IR selected LIRGs at $\bar{z}=$ 0.7 located in the
Hubble Deep Field South was studied by Franceschini et al. (2003) who
computed their stellar masses using a Salpeter IMF and a combination
of single stellar populations with ten different ages, to fit their
UV-optical-NIR spectra of IR luminous galaxies. From their Table 6, a
total of 14 LIRGs possess a spectroscopic redshift between $z$=0.4 and
1.2, and 7 more have a photometric redshift in this range. After
converting Franceschini's values to H$_o$= 75 km s$^{-1}$ Mpc$^{-1}$,
we find a median stellar mass of 7$\times$10$^{10}$ M$_{\odot}$ for
the 14 galaxies with a spectroscopic redshift. Including the less
robust photometric redhistfs only changes this value to
6$\times$10$^{10}$ M$_{\odot}$. Combined with a median D4000$\sim$
1.2, this places the distant LIRGs in a very similar locus as the local
LIRGs (see Fig.~\ref{FIG:masses}).  Their low D4000 values are therefore
also representative of the young stellar population of the burst
superimposed on top of an older stellar population dominating the
stellar mass of the galaxies.

\section{Description of the model}
\label{SEC:simul}
\begin{figure*}
\resizebox{\hsize}{!}{\includegraphics{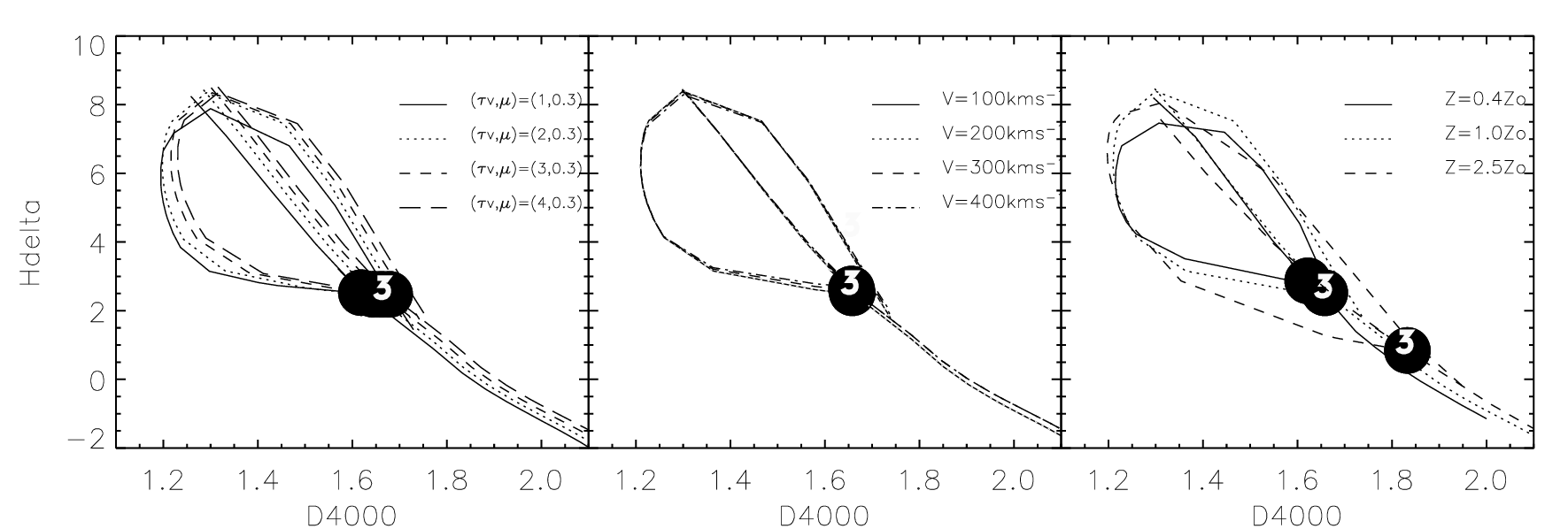}}
\caption{Influence of the attenuation, velocity dispersion and
metallicity for a typical star formation history synthesized with the
GALAXEV code (Bruzual $\&$ Charlot, 2003). In all three plots, the
loops correspond to a burst of star formation which started 3 Gyr (big black circles)
after the galaxy's birth.  a) {\bf{Influence of the attenuation
$\tau_{\rm v}$}} : for a continuous star formation ( $\gamma$=1.0,
Vdisp=200km.$s^{-1}$, $\mu$=0.3, Z=Z$_{\odot}$ ) and a single burst (
$\tau_{\rm B}$=0.1 Gyear and fb=10 $\%$ ); the solid line :
$\tau_{ \rm v}$=1.0. The dotted line : $\tau_{\rm v}$=2.0. The dashed line :
$\tau_{ \rm v}$=3.0. The long dashed line : $\tau_{\rm v}$=4.0. {\bf{b) Influence
of the velocity dispersion Vdisp}} : for a continuous star formation (
$\gamma$=1.0, $\tau_{\rm v}$=3.0, $\mu$=0.3, Z=Z$_{\odot}$) and a single
burst ($\tau_{\rm B}$=0.1 Gyear and fb=10$\%$). The solid line : Vdisp=100
km.s$^{-1}$. The dotted line : Vdisp=200km.s$^{-1}$. The dashed line :
Vdisp=300 Km.$s^{-1}$. {\bf{ c)Influence of the metallicity Z : }}for
a continuous star formation ( $\gamma$=1.0, Vdisp=200km.s$^{-1}$, $\tau_{\rm v}$=3.0 and
$\mu$=0.3 ) and a single burst ($\tau_{\rm B}$=0.1 Gyear and fb=10
$\%$). The solid line : Z=0.5Z$_{\odot}$. The dotted line :
Z=$Z_{\odot}$. The dashed line : Z=2.5Z${\odot}$.  }
\label{FIG:influ}
\end{figure*}
\begin{table}
\begin{center}
\begin{tabular}{|c|c|}
\hline
\multicolumn{2}{|c|}{Continuous star formation} \\
\hline
$t_{\rm form}$ (Gyr)  & 1--7   \\
$\gamma$ (Gyr$^{-1}$)    &   0--3    \\
\hline	     
\multicolumn{2}{|c|}{Metallicity \& velocity dispersion} \\
\hline
Z/Z$_{\odot}$& 0.5--1.5      \\
V$_{\rm disp}$ (km.s$^{-1}$)& 100--300 \\
\hline
\hline
\multicolumn{2}{|c|}{Attenuation} \\
\hline
$\tau_{\rm V}$     & 1--4     \\
$\mu$        & 0.1--1.0      \\
\hline
\multicolumn{2}{|c|}{Burst fraction and duration} \\
\hline
$f_{\rm B}$        & 0-1     \\
$\tau_{\rm B}$ (10$^8$ yr)    & 0.1--10   \\
\hline
\end{tabular}
\end{center}
\caption{Description of the range of values used as priors for the simulations.}
\label{TAB:range}
\end{table}

We used the Bruzual \& Charlot (2003) model to synthesize a series of
200,000 Monte Carlo realizations including various star formation
histories. Although the technique is similar to the one used in Kauffmann
et al. (2003), it presents two differences related to the populations of
galaxies that we are studying here, i.e. dusty starbursts. The priors of
the model were set to include a range of dust attenuations and to
include a larger fraction of starbursting galaxies. We will  discuss
the effects of both modifications in the following.

Each star formation history was modeled with eight parameters:

- the age of the galaxy since its formation, t$_{\rm form}$. We imposed a maximum age of 7 Gyr, which is the age of the universe at the median redshift of the galaxies, ${\bar z}\sim$ 0.7.

- the parameter $\gamma$ for the underlying "continuous" SFR (as opposed to the starburst itself): SFR$^{\rm cont}({\rm t}) \propto e^{-\gamma {\rm{t}}}$

- the SFR of the burst itself, SFR$^{\rm burst}$, is not a parameter but
results from the combination of two parameters: the burst duration,
$\tau$$_{\rm B}$, and the stellar mass fraction produced during the burst,
$f_{\rm B}$. Note that f$_{\rm B}$ computed in the model is the ratio of
  the stellar mass formed during the burst, M$^{\rm burst}_{\star}$, over
the total stellar mass formed through continuous star formation
during t$_{\rm form}$,   M$^{\rm cont}_{\star}$. Here M$^{\rm
cont}_{\star}$ is not corrected for the mass   returned to the
interstellar medium by evolved stars and is therefore larger   than
the actual stellar mass of the galaxy, M$_{\star}$. In the following, we
will   call, f$_{\rm B}^{\rm eff}$, the "effective burst stellar mass
fraction", computed   with M$_{\star}$ instead of M$^{\rm
cont}_{\star}$.

- the dust attenuation parameters, $\mu$ and $\tau_{\rm V}$, defined as in Charlot \& Fall (2000), i.e. a differential attenuation function of the age of the stars (see Eq.~\ref{EQ:ext_law} and Fig.~\ref{FIG:influ}a). Stars younger than $10^7$ years, the typical lifetime of a giant molecular cloud (GMC),
are assumed to be embedded in their dusty parent GMC with an optical
depth $\tau_{\rm V}$. Older stars are
supposed to have escaped their parent GMC and their ambient medium's
optical depth is assumed to be $\mu$ times smaller. This attenuation
law was chosen because it reproduces well the correlation observed for UV selected starbursts of
the FIR over UV ratio with the UV slope, $\beta$, as well as the
L$_{{\rm{H}\alpha}}$ over 
L$_{{\rm{H}\beta}}$ ratio. The
optical depth for stars younger than 0.1 Gyears, $\tau_{\rm V}$, was
set between 1 and 4 based on the observed values for LIRGs (Liang {\it{et  al.}} 2004, Flores {\it{et al.}}, 2004), while $\mu$ was allowed to vary from 0.1 to 1.
\begin{equation}
\begin{tabular}{lll}
$F_{\lambda}^{{\rm observed}}$ ~ = ~ $F_{\lambda}^{{\rm intrinsic}}~\times~e^{-\tau_{\lambda}}$ && \\
$\tau_{\lambda}~=~\tau_{\rm V}~\times~(5500/\lambda)^{0.7}$ & & t$_{\star}$ $<$ 10$^7$ yr \\
$\tau_{\lambda}~=~\mu~\times~\tau_{\rm V}~\times~(5500/\lambda)^{0.7}$ & & ${\rm t_{\star} \geq 10^7 yr}$
\end{tabular}
\label{EQ:ext_law}
\end{equation}

- the velocity dispersion, V$_{\rm disp}$. We allowed a large range of velocity dispersions to test the effect of a broadening of the H$\delta_A$ absorption lines that would mimick intrinsically large width of the absorption lines due to the combination of the internal dynamics of the stars with the mass of the galaxy. The influence of the velocity dispersion on the position of a galaxy in the H$\delta_A$-D4000 diagram was found to be negligible (see Fig.~\ref{FIG:influ}b).

- the metallicity, Z. The influence of the stellar metallicity is stronger than that of the velocity
dispersion but remains smaller than the error bars on the measured $H\delta$ and $D4000$ (see Fig.~\ref{FIG:influ}c).

Table~\ref{TAB:range} summarizes the range over which those eight parameters were chosen, the so-called priors, which represent the key adjustment when using a Bayesian statistics. The goal of these simulations is not to constrain all eight parameters and indeed some of them produce competitive effects or cannot be disentangled:

- a galaxy with a given age and $\gamma$ parameter for its SFR will be located at the same position in the H8-D4000 diagram than a younger galaxy with a larger value for $\gamma$.

- the two parameters used in the attenuation law, $\mu$ and $\tau_{\rm V}$. Fig.~\ref{FIG:influ} shows that galaxies are shifted towards the upper-right of the H$\delta_A$-D4000 (or equivalently H8-D4000) diagram with increasing dust attenuation. This is an important point to note when comparing dusty galaxies to field galaxies: even though LIRGs overlap with a fraction of SDSS field galaxies in the Fig.~\ref{FIG:irascam}, their recent star formation history is interpreted differently if their dust attenuation is taken into account. 

Instead, we will use a bayesian approach such as the one detailed in
Kauffmann et al. (2003) to derive the following characteristics of the
starbursts:

\begin{enumerate}
\item the Scalo parameter, i.e. the ratio of 
present to past-averaged star formation rate. This parameter defines the relative intensity of the present starburst.
\item the burst duration, $\tau_{\rm B}$.
\item the last burst stellar mass fraction, $f_{\rm B}$ (and $f_{\rm B}^{\rm eff}$).
\end{enumerate}

The bayesian statistics approach consists in determining a probability
distribution function (PDF) for any one of these parameters, which
consists in an histogram of the number of Monte Carlo realizations
weighted by the probability function $exp(-\chi^2/2)$, where $\chi^2$
is defined in Eq.~\ref{EQ:chi2}. The convergence of the technique is
reached if the PDF is peaked and its shape provides the precision of
the determination.

\begin{equation}
\chi2~=
\left(\frac{H\delta^{\rm obs}-H\delta^{\rm sim}}{\sigma(H\delta^{\rm obs})}\right)^2~+~\left(\frac{D4000^{\rm obs}-D4000^{\rm sim}}{\sigma(D4000^{\rm obs})}\right)^2 
\label{EQ:chi2}
\end{equation}

\section{Results}
\subsection{Signature of the presence of a starburst in the distant LIRGs in the H8-D4000 diagram}
\begin{figure*}
\resizebox{\hsize}{!}{\includegraphics{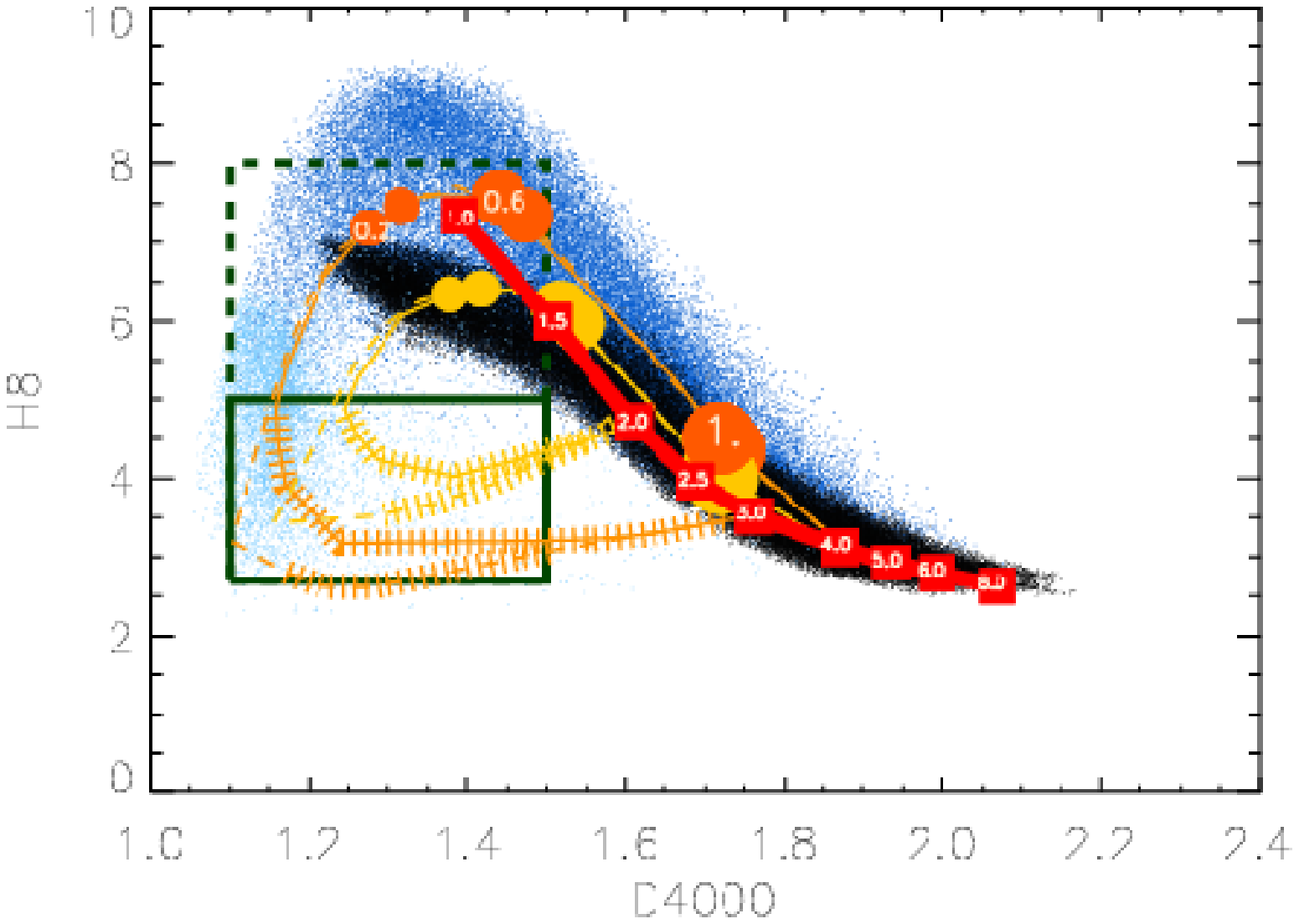}}
\caption{Location of the 200,000 Monte Carlo realizations using the parameters of SIM1 (see table~\ref{TAB:range}) in the H8 versus 4000\,\AA (D4000) break diagram. {\bf Light blue points (lower-left):} starbursting galaxies. {\bf Dark blue points (upper part):}
post-starburst galaxies (galaxies having experienced a recent starburst which ended less than 2 Gyr ago). {\bf Black points:} galaxies whith a continuous
star formation in the two past Gyears. {\bf Bold red line:} track
followed by an individual galaxy with continuous star formation
(numbers in squares= age in Gyr) generated with GALAXEV ($\gamma$=2.,
$\mu$=0.3, $\tau_{\rm V}$=3.0,V$_{\rm disp}$=200 km.s$^{-1}$). {\bf
Orange loops:} effect of starbursts of 5$\times$10$^7$ (dashed line)
and 10$^8$ (plain line) years starting after 2 Gyr (light orange) and
3 Gyr (dark orange) of continuous star formation. The line is hatched
during the starburst phase.  The size of the orange dots is
proportional to the time counted in 10$^8$ years units after the
beginning of the burst. {\bf The dark rectangle with dashed contours:}
area where the distant LIRGS are located. {\bf The dark rectangle with
a solid line:} area where are lying 75 $\%$ of the sample (see
Fig.~\ref{FIG:plot_avec_cam} for the position of the individual
galaxies with error bars). 
{\bf Triangles:}
galaxies corrected for the underlying nebular emission line in H8 as
derived from the observed lower order Balmer lines corrected
for attenuation from the Balmer lines ratio. 
{\bf Stars:} galaxies with
negligible emission Balmer line in emission, hence uncorrected for H8 emission. 
{\bf Squares:} corrected for
the underlying H8 emission line using only one Balmer emission line
available with dust attenuation computed from the IR (from
the ratio of SFR(IR) over SFR(H$\beta$)). 
{\bf Circles:} no correction
for the underlying potential emission line because no information is 
available on the Balmer emission lines.
 }
\label{FIG:plot_ss_cam}
\end{figure*}
\begin{figure*}
\resizebox{\hsize}{!}{\includegraphics{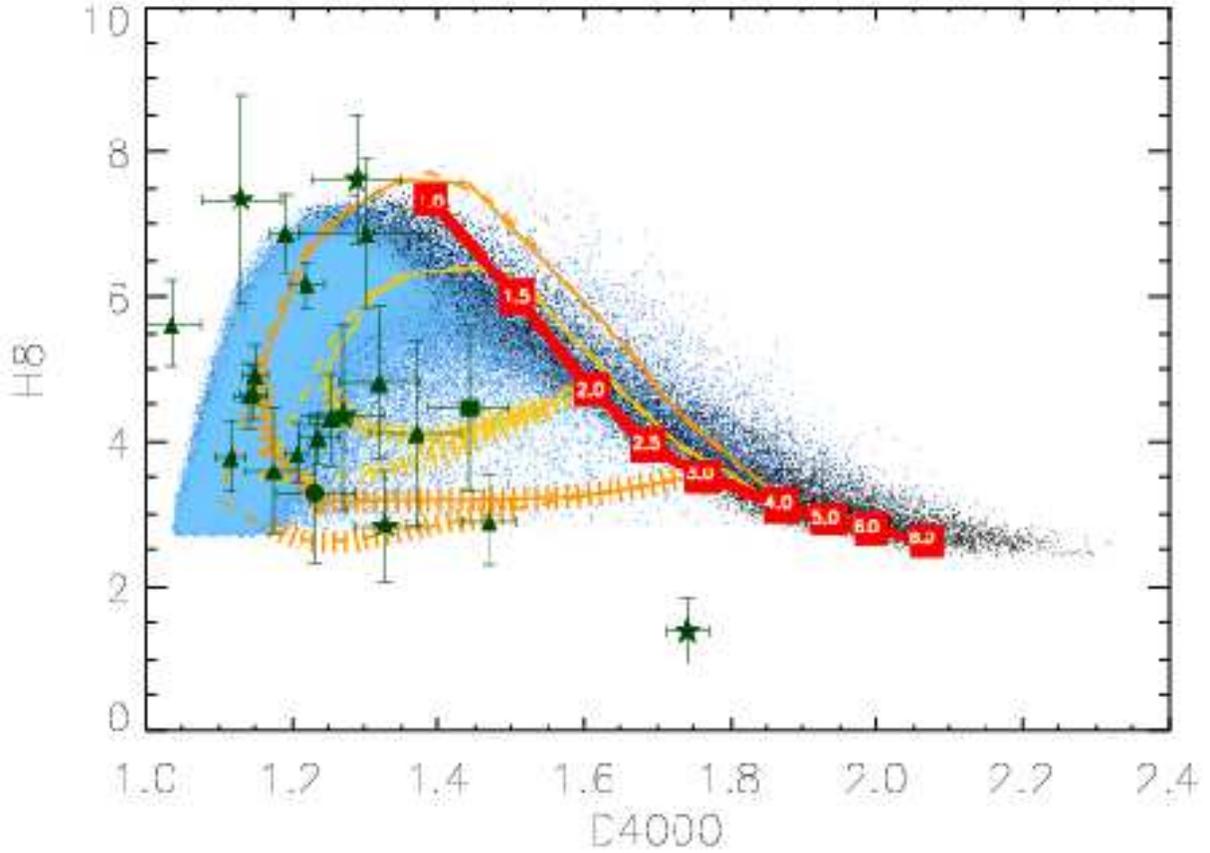}}
\caption{Same as in Fig.~\ref{FIG:plot_ss_cam} for the SIM2
Monte Carlo simulation of 200 000 model galaxies. The distant LIRGs
are represented with green points with error bars.  
}
\label{FIG:plot_avec_cam}
\end{figure*}
\begin{figure*}
\resizebox{\hsize}{!}{\includegraphics{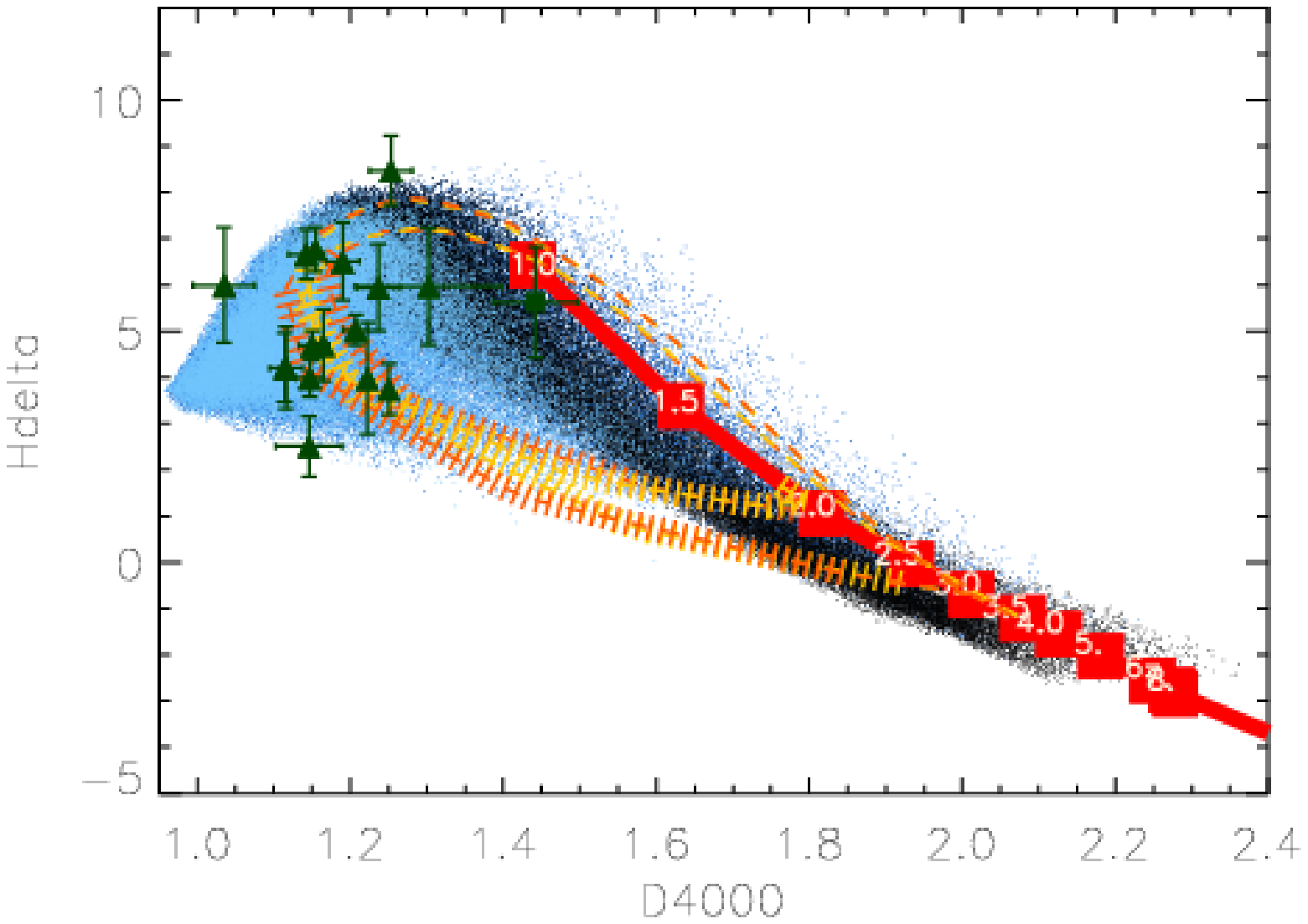}}
\caption{
Location of the 200,000 Monte Carlo realizations using the parameters of SIM2 (see table~\ref{TAB:range}) in the H$\delta_A$ versus 4000\,\AA (D4000) break diagram. {\bf Light blue points (lower-left):} starbursting galaxies. {\bf Dark blue points (upper part):}
post-starburst galaxies (galaxies having experienced a recent starburst which ended less than 2 Gyr ago). {\bf Black  points:} galaxies whith a continuous
star formation in the two past Gyears. {\bf Bold red line:} track
followed by an individual galaxy with continuous star formation
(numbers in squares= age in Gyr) generated with GALAXEV ($\gamma$=4.,
$\mu$=0.3, $\tau_{\rm V}$=3.0,V$_{\rm disp}$=200 km.$s^{-1}$). {\bf
Orange loops:} effect of starbursts of 5$\times$10$^7$ (dashed line)
and 10$^8$ (plain line) years starting after 2 Gyr (light orange) and
3 Gyr (dark orange) of continuous star formation. The line is hatched
during the starburst phase.  The size of the orange dots is
proportional to the time counted in 10$^8$ years units after the
beginning of the burst. 
{\bf Triangles:}
galaxies corrected for the underlying nebular emission line in H$\delta_A$ as
derived from the observed lower order Balmer lines corrected
for attenuation from the Balmer lines ratio. 
{\bf Stars:} galaxies with
negligible emission Balmer line in emission, hence uncorrected for H$\delta_A$ emission. 
{\bf Squares:} corrected for
the underlying H$\delta_A$ emission line using only one Balmer emission line
available with dust attenuation computed from the IR (from
the ratio of SFR(IR) over SFR(H$\beta$)). 
{\bf Circles:} no correction
for the underlying potential emission line because no information is 
available on the Balmer emission lines.
}
\label{FIG:Hd}
\end{figure*}
We generated three series of simulations with the same priors but varying the percentage of galaxies experiencing a starburst at the age of the simulated spectrum. In a first simulation, SIM1, only 20\,\% of the Monte Carlo realizations include ongoing starburst. The positions in the H8-D4000 diagram of the 200,000 Monte Carlo realizations of SIM1 are illustrated with small dots in Fig.~\ref{FIG:plot_ss_cam}. The light blue points on the lower-left part of the figure illustrate the position of galaxies with an ongoing burst of star formation. The upper dark blue points correspond to post-starbursts, i.e. galaxies having experienced a recent starburst which ended less than 2 Gyr ago, and dark points to galaxies with continuous star formation or at least no burst during the past 2 Gyears.  The lines illustrate the tracks followed by an individual galaxy during its lifetime: a galaxy with purely continuous star formation will follow the bold red line, where the numbers mark its age in Gyr ($\gamma$=2, $\mu$=0.3, $\tau_{\rm V}$=3.0,V$_{\rm disp}$=200 km.$s^{-1}$). If this galaxy was to experiment a
burst of star formation when it is 3 Gyrs old, then it would
follow one of the loops in light orange towards the left of the diagram for a total duration of 2 Gyrs and then go back to the track of a galaxy with pure continuous star formation, having lost the memory of its past starburst. During the loop, D4000 is at first the most affected by the presence
of young stars, moving towards the left of the diagram, then it is H8 which increases moving
upwards in the diagram, before the galaxy slowly goes back to the
continuous regime. The first part of the loop is marked with small
lines indicating that the galaxy is still bursting, while the
remaining part of the loop corresponds to the post-starburst
regime. We have plotted a loop starting at an age of 2 Gyears (light orange) and another starting at 3 Gyears (dark orange). In each case we computed the track of $\tau_B$=5$\times$10$^7$ years (dashed lines) and 10$^8$years (plain line) starburst with a fraction of $f_{\rm B}$=15\,\%.

75\,\% of the distant LIRGs are located within the solid rectangle in Fig.~\ref{FIG:plot_ss_cam}, hence at the location of galaxies experiencing a starburst in the simulation. The remaining 25\,\% of the distant LIRGs lie in the dashed rectangle which includes the region of post-starbursts but also galaxies with successive or longer starbursts. Hence, the results of the simulation from SIM1 simulation show that the observed galaxies are mainly coherent with being starbursting galaxies.

As a second step, we designed a new simulation, with 80\,\% of the galaxies experiencing an ongoing starburst (SIM2), to better sample the locus of the distant LIRGs and therefore better study the properties of the starbursts themselves. A third and last simulation (SIM3) was generated to quantify the probability that the distant LIRGs experienced a previous starburst during the last 2 Gyr. Half of the galaxies in SIM3, i.e. 100,000 galaxies, have experienced a previous starburst during the last 2 Gyr, which ended before the onset of the ongoing starburst.

The individual positions of the distant LIRGs are compared to the SIM2
realizations in Fig.~\ref{FIG:plot_avec_cam} and Fig~\ref{FIG:Hd} in
the H8-D4000 and H$\delta_A$-D4000 diagrams respectively. 
UDSR09, which is lying at the bottom of the simulation is a clear outlyer. Liang et al (2005) studied carrefully this object and its optical spectrum shows strong metal absorption lines such as Ca H K, G-band, Mg H, Na D lines but weak Balmer absorption lines. The X-ray emission of this object, in addition  is mainly associated to an AGN. As a consequence, an AGN can partially contribute to the MIR luminosity. 
Note that
the observed galaxies are not identical in both figures neither in
numbers nor in identity because both H8 and H$\delta_A$ cannot be both
measured for all individual galaxies.  The dots correspond to the same
simulated galaxies in both figures and in both cases. Note that the
observed LIRGs do lie below the continuous star formation regime in
both figures, although in slightly different locations. These
differences will be discussed in the subsection ~\ref{test} when we will present
the resulting PDFs. Note also that the distant LIRGs are distributed
in two populations in Fig.~\ref{FIG:plot_avec_cam}, one located at the
bottom of the diagram and a second at the upper-left. The second
population corresponds to simulated galaxies which experienced a
succession of two starbursts during the last 2 Gyrs. Such histories
are expected in the framework of major mergers of spiral galaxies with
several encounters between two galaxies.

\subsection{Determination of the Scalo parameter, SFR/$<$SFR$>$}
\begin{table*}
\begin{center}
\begin{tabular}{|l|l|l|l|l|}
\hline
\hline
id      &   SFR/$<$SFR$>$                & ($\tau_{\rm B}$) [($\tau_{\rm B}^-$),($\tau_{\rm B}^+$)]   &  $f_{\rm B}^{\rm }$ [$f_{\rm B}^{\rm ~-}$,$f_{\rm B}^{\rm ~+}$]&$f_{\rm B}^{\rm eff}$ [$f_{\rm B}^{\rm eff~-}$,$f_{\rm B}^{\rm eff~+}$] \\ 
        &    $[$68\,\%$]$        &$\times$0.1 Gyr $[$68\,\%$]$                                         &(\%) $[$68\,\%$]$&(\%) $[$68\,\%$]$\\
\hline
\hline
UDSF07   &      6 [      3,      24  ] &	      ~~--~~   & 4 [ 3,17]  &    5     [3,25  ] \\
UDSF16  &      ~~--~~		    &	      0.8   [   0.3   ,   3.0	   ]& 3[2,5 ]   &     4    [3,7  ] \\
UDSF17  &      2 [      1,      5    ] &	      4    [   1.5   ,  6	   ]& ~~--~~       &   ~~--~~             \\
UDSF18  &      2 [      1,      5    ] &	      1   [   0.5   ,  3	   ]& 4 [ 3,13]  &    5     [3,17  ] \\
UDSF19  &      ~~--~~ &	      1.6   [   0.4   ,   4.0	   ]& 2 [1,23 ]  &    3     [2,10  ] \\
UDSF31  &      ~~--~~		    & ~~--~~   & ~~--~~       &    ~~--~~       \\
\hline
UDSR08   &      6 [      3,      27    ] &	      0.4   [   0.2   ,  1.3	   ]& 3 [ 2,16]  &    5    [3,26  ] \\
UDSR10  &      ~~--~~		    &  ~~--~~  & ~~--~~        &   ~~--~~      \\
UDSR14  &      4 [      2,      13   ] &	      0.3   [   0.1   ,   0.8	   ]& 4 [2,9  ]  &     5    [3,14  ] \\
UDSR20  &      ~~--~~		    & 5   [   1   , 8         ]& ~~--~~       &     ~~--~~       \\
UDSR23  &      3 [      1,      10  ] &	      0.8   [   0.3   ,   1.6	   ]& 6 [4,9  ]  &    7     [5,19  ] \\
\hline
CFRS02   &      ~~--~~		    &~~--~~    & ~~--~~       &     ~~--~~ \\
CFRS06   &      2 [      1,      4    ] &	      2.    [   1   ,   4	   ]& 2 [1,9  ]  &     7    [5,19  ] \\
CFRS08   &      ~~--~~		    &	    ~~--~~    & ~~--~~       &     ~~--~~ \\
CFRS10  &      10 [      5,      36   ] &	      1    [   0.2   ,   2.0	   ]& 13 [7,30]  &      14   [7,47  ] \\
CFRS11  &      3 [      2,      15   ] &	      ~~--~~   & 3 [2,5  ]  &      4   [3,9  ] \\
CFRS14  &      3 [      2,      6    ] &	      3   [   1   ,   6	   ]& ~~--~~       &      ~~--~~\\
\hline
\hline
\hline
\end{tabular}
\end{center}
\caption{Results concerning the burst duration, the Scalo ratio
SFR/$<$SFR$>$, and the  burst fraction ($f_{\rm B}$) when using the H$\delta_A$ data.}
\label{TAB:data}
\end{table*}

\begin{table*}
\begin{center}
\begin{tabular}{|l|l|l|l|l|}
\hline
\hline
id & SFR/$<$SFR$>$& $\tau_{\rm B}$ [$\tau_{\rm B}^-$,$\tau_{\rm B}^+$] &$f_{\rm B}$ [$f_{\rm B}^-$,$f_{\rm B}^+$]& $f_{\rm B}^{\rm eff}$ [$f_{\rm B}^{\rm eff~-}$,$f_{\rm B}^{\rm eff~+}$] \\ 
   & (\%) $[$68\,\%$]$ & $\times$0.1 Gyr $[$68\,\%$]$&(\%) $[$68\,\%$]$&(\%) $[$68\,\%$]$\\
\hline
\hline  
UDSF06   & ~~--~~             & 2  [1,6]   &      3  [2 ,19 ]     &   3 [2,22]         \\
UDSF07   & 4 [2,13]           & 0.2 [0.1,0.9]       &      ~~--~~                &   ~~--~~                \\
UDSF12   & ~~--~~             &  $\geq$8.9          &      ~~--~~		 &    ~~--~~                \\
UDSF17   &  ~~--~~            & 5  [1,8]           &      ~~--~~		 &   ~~--~~                 \\
UDSF19   & ~~--~~             & ~--~                &      2  [1 ,15 ]       &   3 [2 19  ]      \\
UDSF28   & 4  [2,8]           & ~--~                &      ~~--~~		 &    ~~--~~                \\
UDSF29b  & ~~--~~             & ~--~                &      ~~--~~		 &    ~~--~~                \\
UDSF20   & ~2~ [1,5]  & ~--~                &      ~~--~~		 &   ~~--~~                 \\
UDSF31   & ~~--~~             & ~--~                &      ~~--~~		 &   ~~--~~                 \\
\hline
UDSR08   & 7~ [3,30]           & 0.4  [0.2,0.7]      &      ~~--~~		 &   ~~--~~                 \\
UDSR09   &  ~~--~~            & 0.4  [0.2,0.9]   &      ~~--~~		 &   ~~--~~                 \\
UDSR10   &   ~~--~~           & 1  [0.3,2.2]      &      3  [2 ,13 ]        &      3 [3,15]          \\
UDSR14   & ~5~ [2,12]         & 0.8  [0.3,1.7]      &      ~~--~~		 &      ~~--~~              \\
UDSR20   &  ~~--~~            & 7  [3,8]          &      ~~--~~		 &      ~~--~~              \\
UDSR23   & ~3~ [2,7]           & 1  [1,3]      &      ~~--~~		 &   ~~--~~                 \\
\hline
CFRS02   & ~~--~~             & $\geq$3.1          &      ~~--~~		 &   ~~--~~                 \\
CFRS06   & ~3~ [1,7]           & 0.2  [0.1,0.7]      &      4  [3 ,13 ]     &   4 [3,16]         \\
CFRS08   & ~4~ [2,10] & 0.4  [0.1,1.0]     &      3  [2 ,11]          &   4 [3,14]          \\
CFRS10   & ~8~ [5,14] & 1.0  [0.6,2.0]     &      ~~--~~		 &   ~~--~~                 \\
CFRS14   & ~4~ [2,11]          & 0.6  [0.3,1.6]      &      ~~--~~		 &   ~~--~~                 \\
CFRS16   & ~7~ [3,23]   & $\leq$7.85          &      4  [2 ,18 ]         &   5 [3,21]         \\
CFRS29   & ~3~ [2,~9]     & 2.5  [1,5]   &      ~~--~~		 &   ~~--~~                 \\

\hline
\hline
\end{tabular}
\end{center}
\caption{Results concerning the burst duration, the Scalo ratio
SFR/$<$SFR$>$, and the  burst fraction ($f_{\rm B}$) when using the H8 data. }
\label{TAB:dataH8}
\end{table*}

\begin{figure}
\resizebox{\hsize}{!}{\includegraphics{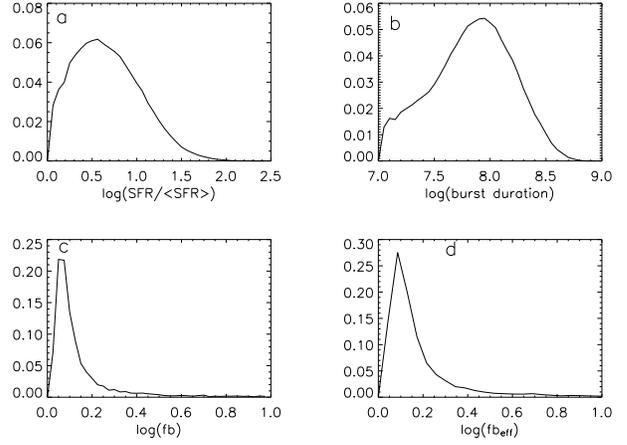}}
\caption{Example of PDF obtained for 1 galaxy   :
a) The Scalo parameter (subsection 6.2).
b  Burst duration (subsection 6.2).
c) Burst fraction (subsection 6.2).
d) Effective burst fraction (subsection 6.2).}
\label{FIG:scalo}
\end{figure}

\begin{figure}
\resizebox{\hsize}{!}{\includegraphics{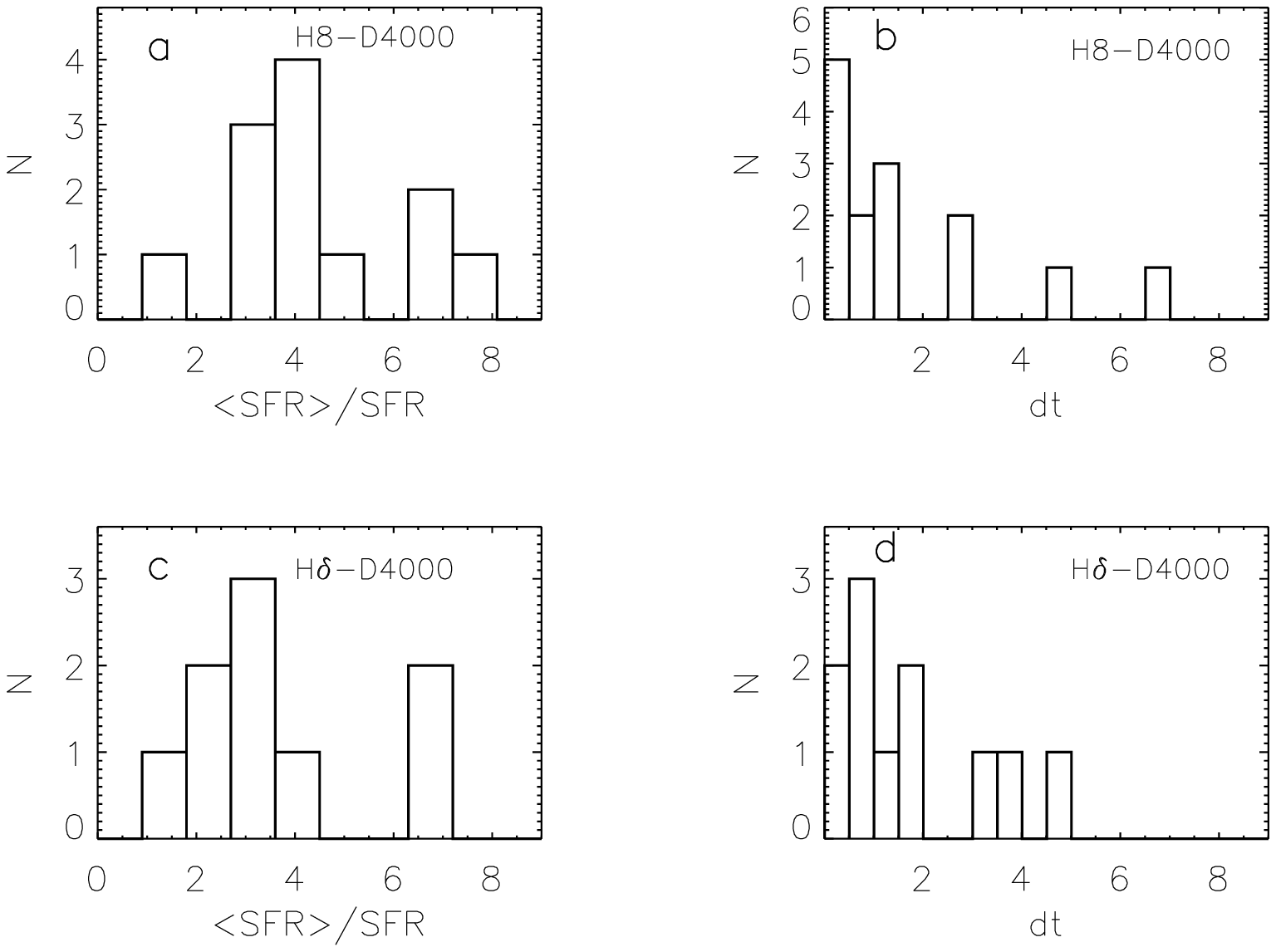}}
\caption{Distribution of the Scalo parameter and the burst duration. 
a) The Scalo parameter in the H8-D4000 diagram. 
b) The burst duration in the H8-D4000 diagram.
c) The Scalo parameter in the H$\delta_A$-D4000 diagram. 
d) The burst duration in the H$\delta_A$-D4000 diagram.}
\label{FIG:result1}
\end{figure}

The PDF obtained for the Scalo parameter of a typical galaxy with SFR/$<$SFR$>\sim$
2.7$_{-1.0}^{+7.3}$ (1-$\sigma$) is represented in Fig.~\ref{FIG:scalo}a. In about 60\,\% of the distant LIRGs, the PDF converge towards a determination of the Scalo parameter. For 10 out of 17 galaxies for H$\delta_A$ (see Table~\ref{TAB:data}), we compute a median value of 
SFR/$<$SFR$>=~3.0^{+3.5}_{-0.6}$ (1-$\sigma$). For 12 out of 22 for H8 (see Table~\ref{TAB:dataH8}) we find SFR/$<$SFR$>$= 4$^{+1}_{-1}$ (1-$\sigma$). These two results are consistent with each other, as illustrated by the Figs.~\ref{FIG:result1}a and c. The fact that the H8-D4000 diagram provides the tightest constraint on the Scalo parameter is due to the fact that the sky background is lower in the bluer wavelength range of H8 than in H$\delta_A$. We will therefore use that range of values for the Scalo parameter in the next set of computations. 
The fact that the LIRGs are found to produce stars at a rate about four times larger than their averaged past SFR confirms that they are experiencing a major phase of star formation in their lifetime. This result is consistent with the large $L_{\rm IR}$ and SFR(IR).

As noted above, the H8-D4000 diagram provides the tightest constraint on the Scalo parameter with SFR/$<$SFR$>$= 4$^{+1.0}_{-0.9}$. Combined with the ongoing SFR measured from the MIR emission for the distant LIRGs (quoted in Table~\ref{TAB:echantillon}) for individual galaxies and with a median value of:   
$\overline{{\rm SFR}}=$ 52 $^{+34}_{-33}$ M$_{\odot}$ yr$^{-1}$
, the Scalo parameter allow us to compute the averaged past SFR of the distant LIRGs : $<$SFR$>\sim$ 13$\pm$3 M$_{\odot}$ yr$^{-1}$. 

If we assume that the progenitors of the distant LIRGs formed stars
at a constant rate equal to the averaged past SFR, a median stellar
mass of 7$\times$10$^{10}$ M$_{\odot}$ (see Sect.~\ref{SEC:comp}) is
assembled in about 5.5$\pm$1.5 Gyr (light-weighted age), implying that
the first dominant stellar populations formed at a redshift $z\sim$
4.5$\pm$1.5.
 
\subsection{Determination of the burst duration }
In the H$\delta_A$-D4000 diagram, the burst duration, $\tau_{\rm B}$, can be constrained for 11 galaxies (Table~\ref{TAB:data}) for which we obtain a median of $\tau_{\rm B}=1.3^{+1.9}_{-0.5}\times10^8$ years, e.g. Fig.~\ref{FIG:scalo}b. Using H8, the technique converges for 14 galaxies to $\tau_{\rm B}=1.0_{-0.6}^{+1.6}\times10^8$ years. Here again, as for the Scalo parameter in the previous section, we find a consistent result using both H8 and H$\delta_A$. Both indicate that the bursts are short-lived over about 0.1 Gyr. The large dispersion between the various galaxies may result from different initial gas mass fractions or triggering mechanisms for the burst.

\subsection{Determination of the burst stellar mass fraction}
The simulations converge to a value of $f_{\rm B}$ for only 6/22 galaxies
in the H8-D4000 diagram and 10/17 in the H$\delta_A$-D4000 (see the example of UDSR 23 in
Fig.~\ref{FIG:scalo}c). The difficulty to determine the burst stellar mass fraction is not surprising. It results from the fact that as soon as the young stellar population starts dominating the spectrum of a galaxy, it becomes nearly impossible to measure with good precision the fraction of the light due to the underlying older population. Indeed, the PDFs converge only for galaxies for which the fraction of young stars remains low (see Tables~\ref{TAB:data},\ref{TAB:dataH8}). 

However, we can still derive the effective burst stellar mass fraction, $f_{\rm B}^{\rm eff}$, which is equal to the mass of stars produced during the burst, i.e. $\tau_{\rm B}\times$ SFR, divided by the total stellar mass of the galaxy, i.e. $t_{\rm form}\times<$SFR$>$. Since $\tau_{\rm B}=1.0_{-0.6}^{+1.6}\times10^8$ years and SFR=52 $^{+34}_{-33}$ M$_{\odot}$ yr$^{-1}$, we obtain an effective burst stellar mass fraction of $f_{\rm B}^{\rm eff}=$ 10$\pm$9\,\%. Note that a burst producing 10\,\% of a 7$\times$10$^{10}$ M$_{\odot}$ galaxy, will convert 7$\times$10$^{9}$ M$_{\odot}$ of molecular gas into stars, which is consistent with the mass of molecular gas observed in local LIRGs and ULIRGs (Sanders \& Mirabel 1996).

\subsection{Testing the "multiple burst" scenario}
\label{test}
\begin{figure}
\resizebox{\hsize}{!}{\includegraphics{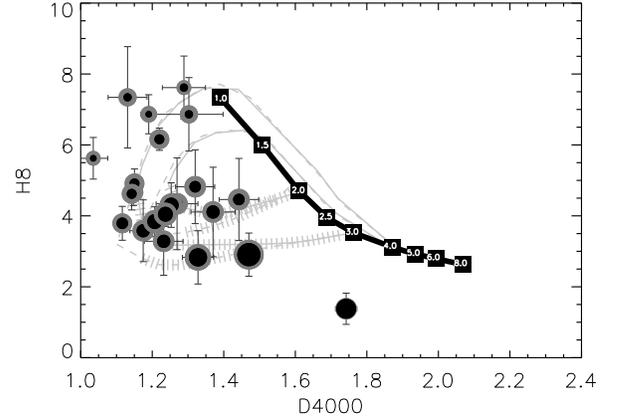}}
\caption{The lines are the same as in Fig.~\ref{FIG:plot_ss_cam} and
in Fig.~\ref{FIG:plot_avec_cam}. The black circle are proportional to
$t_{20}$ defined before, while the grey one to $t_{50}$.}
\label{FIG:intb}
\end{figure}

\begin{figure}
\resizebox{\hsize}{!}{\includegraphics{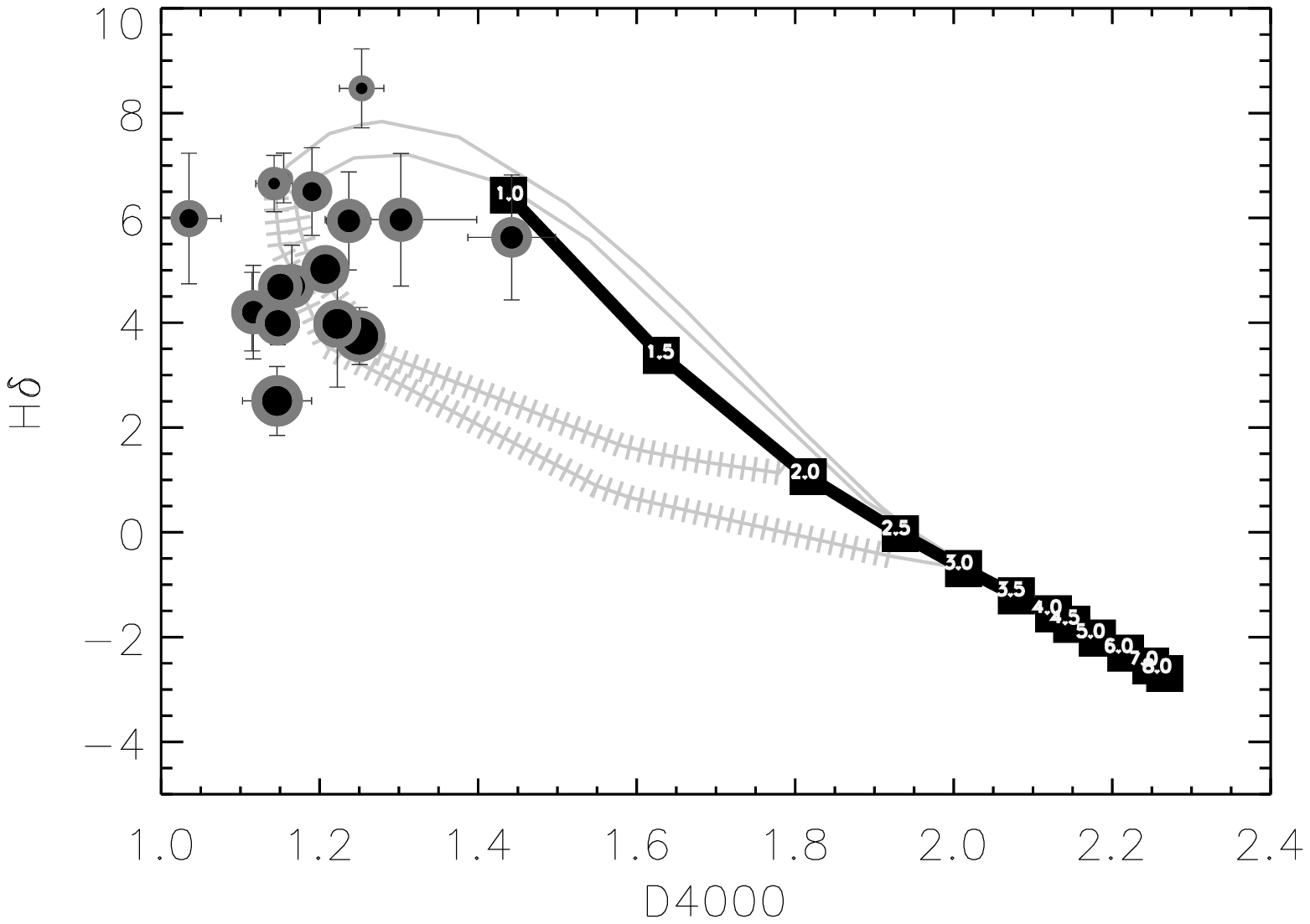}}
\caption{The lines are the same as in Fig.~\ref{FIG:plot_ss_cam} and
in Fig.~\ref{FIG:plot_avec_cam}. The black circle are proportional to
$t_{20}$ defined before, while the grey one to $t_{50}$.}
\label{FIG:intbHd}
\end{figure}

\begin{table}
\begin{center}
\begin{tabular}{l|l|l}
\hline
\hline
id      &$t_{20}$       &$t_{50}$         \\
        &$10^8years$  &$10^8years$  \\
(1)&(2)&(3)\\
\hline
\hline

6.udsf  & 7.0  &  13.0 \\
7.udsf  & 8.0  &  13.0 \\
12.udsf & 5.0  &  9.0  \\
17.udsf & 6.0  &  12.0 \\
19.udsf & 7.0  &  13.0 \\
28.udsf & 14.0 & 18.0 \\
29.udsf & 8.0  &  13.0 \\
20.udsf & 11.0 & 16.0 \\
31.udsf & 7.0  &  13.0 \\
\hline
8.udsr  & 7.0  &  12.0 \\
10.udsr & 9.0  &  14.0 \\
14.udsr & 7.0  &  12.0 \\
20.udsr & 4.0  &  10.0 \\
23.udsr & 6.0  &  12.0 \\
9.udsr  & 12.0 & 13.0 \\
\hline
2.cfrs  & 5.0  &  11.0 \\
6.cfrs  & 9.0  &  14.0 \\
8.cfrs  & 9.0  &  14.0 \\
10.cfrs & 4.0  &  9.0  \\
14.cfrs & 6.0  &  12.0 \\
16.cfrs & 8.0  &  13.0 \\
29.cfrs & 5.0  &  11.0 \\
\hline
\end{tabular}
\end{center}
\caption{Values of $t_{20}$ (black circle) and $t_{50}$ (grey circle) obtained for the 22 galaxies in the H8-D4000 diagram, to quantify the probability of two successive starbursts (see text).}
\label{TAB:intb}
\end{table}
\begin{table}
\begin{center}
\begin{tabular}{l|l|l}
\hline
\hline
id      &$t_{20}$       &$t_{50}$         \\
        &$10^8years$  &$10^8years$  \\
(1)&(2)&(3)\\
\hline
\hline
       7.udsf & 6.0  & 11.0 \\
      16.udsf & 10.0 & 14.0 \\
      17.udsf & 3.0  &  5.0 \\
      18.udsf & 7.0  & 12.0 \\
      19.udsf & 8.0  & 13.0 \\
      31.udsf & 6.0  & 11.0 \\
      \hline		   	 
       8.udsr & 6.0  & 12.0 \\
      10.udsr & 3.0  &  7.0 \\
      14.udsr & 7.0  & 12.0 \\
      20.udsr & 5.0  & 11.0 \\
      23.udsr & 7.0  & 12.0 \\
      \hline 		   	 
      2.cfrs  & 6.0  & 12.0 \\
       6.cfrs & 8.0  & 13.0 \\
       8.cfrs & 6.0  & 12.0 \\
      10.cfrs & 5.0  & 10.0 \\
      11.cfrs & 8.0  & 14.0 \\
      14.cfrs & 3.0  &  9.0 \\
\hline
\end{tabular}
\end{center}
\caption{Values of $t_{20}$ (black circle) and $t_{50}$ (grey circle) obtained for the 17 galaxies in the H$\delta_A$-D4000 diagram, to quantify the probability of two successive starbursts (see text).}
\label{TAB:intbHd}
\end{table}

A sub-sample of the distant LIRGs lie on the upper-left part of the
H8 and H$\delta_A$ versus D4000 diagrams
(Figs.~\ref{FIG:plot_avec_cam},~\ref{FIG:Hd}).
This region is populated by Monte Carlo realizations of galaxies for which the ongoing burst of star formation was superimposed on a previous burst which ended less than 2 Gyr ago. In order to quantify the probability of such an occurence, we generated a third simulation, SIM3, in which half of the realizations experienced two successive bursts during the last 2 Gyr. We then computed for each galaxy, using the same bayesian statistics, a probability for various possible durations between the two bursts. Two representative numbers are assigned for each galaxy in Tables~\ref{TAB:intb} and \ref{TAB:intbHd}: $t_{20}$ and $t_{50}$. These values correspond to lookback times associated with a 20 and 50\,\% chance of finding a previous burst which ended $t_{20}$ and $t_{50}$ Gyr before the onset of the ongoing one. For example, in the case of UDSR20, we obtain from the H8-D4000 diagram (see Table~\ref{TAB:intb}) $t_{20}=$ 0.4 Gyr and $t_{50}=$ 1 Gyr, which implies that there is a 50\,\% chance that a previous starburst occured 1 Gyr before the onset of the present one and a 20\,\% chance that the delay was only a 0.4 Gyr. Hence UDSR20 is a good candidate for two successive bursts. We did not consider higher probabilities or longer timescales because of the limited constraints that we can set on those parameters and because after about 1.5 Gyr, the memory of the previous burst is lost with this technique.

Local LIRGs and ULIRGs are known to be predominantly triggered by
major mergers (Borne et al. 1999, Sanders \& Mirabel 1996) and
numerical simulations of such mergers predict that the two galaxies
cross each other several times, potentially inducing a series of
bursts separated by a few tens of million years (Mihos \& Hernquist
1996).  However, in distant LIRGs, the probability that a previous
starburst occured less than 0.5 Gyr ago is nearly always lower than
20\,\%. This result must be considered together with recent
evaluations of the morphological properties of distants LIRGs which
also suggest that most of them are not produced in major mergers (Bell
et al. 2005, Zheng et al. 2004, Elbaz \& Cesarsky 2004). Bell et
al. (2005) suggested that distant LIRGs could either be non-triggered
phases in isolated spirals with larger gas masses, possibly
experiencing some infall, or minor mergers, where the dwarf galaxy
responsible is not detected. Our determination of a burst duration of
0.1 Gyr and a Scalo parameter of 4, seems to rule out the possibility
that distant LIRGs are isolated spirals forming stars at a constant
rate over a long duration. The starbursts may instead be triggered by
tidal effects and minor mergers in regions of the universe where the
local density of galaxies is enhanced as suggested by Elbaz \&
Cesarsky (2003) or by the infall of intergalactic gas (Combes, 2005). Further kinematical studies of distant galaxies (see Flores et al, 2006; Puech et al, 2006)
will help to distinguish between the various scenarios (mergers, gas infall) discussed here."

\section{Discussion and conclusions}
We have analyzed the star formation history of a sample of 25 distant LIRGs ($\bar{z}=$ 0.7) that we derived from their stellar spectra (Balmer absorption lines and 4000\,\AA\,break). The high order Balmer absorption line H8 and the H$\delta_A$ line provide consistent results although H8 is more adapted to distant galaxies being located in a bluer region of the spectrum, less affected by sky emission lines and which is measured for more distant objects than H$\delta_A$. Variations at the 30\,\% level between the burst parameters obtained using one or the other indicator suggest that some effects, such as abundance ratios which might affect the pseudo-continua surrounding the H$\delta_A$ line, should be taken into account to improve the models (see Thomas, Maraston \& Korn 2004, Korn, Maraston \& Thomas 2005).

The comparison of distant LIRGs, selected from ISOCAM and MIPS, onboard ISO and Spitzer, to local LIRGs, selected from IRAS and the SDSS, shows that both populations present similar spectral features and therefore suggests that they are experiencing comparable starburst phases. The fact that half of the local LIRGs present D4000 values larger than the maximum D4000 of distant LIRGs indicates that the dominant non bursting stellar population is younger for distant LIRGs, as expected.

The first important result of this study is the identification of an optical signature for the presence of a starburst in these galaxies, in spite of their large dust attenuation. While continuous star formation follows a line along decreasing Balmer EW and increasing D4000, a burst superimposed on this population produces a loop that first decrease D4000 and then increase the Balmer line EW. However, after about 1.5 Gyr, the memory of the burst is lost and the galaxy behaves like others which did not experience a starburst. As result, we are limited to study only the averaged past star formation history for lookback times shorter than 1.5 Gyr. The bursts characteristics were derived from probability distribution functions (PDF) using a bayesian statistics as in Kauffmann et al. (2003). 

The median ratio of present over averaged SFR, the so-called Scalo parameter, for the distant LIRGs is SFR/$<$SFR$>=$ 4$\pm$1 (we used the H8 line for which a larger sample of galaxies is available and the PDFs present a sharper peak), which indicates that these galaxies are experiencing an atypically intense phase of star formation in their lifetime. A median SFR of 52$^{+34}_{-33}$ M$_{\odot}$ yr$^{-1}$ for the ongoing starbursts was derived from their MIR luminosities. Hence, their mean SFR averaged over their lifetime is $<$SFR$>=$ 13$\pm$3 M$_{\odot}$ yr$^{-1}$. Knowing the median stellar mass for LIRGs of equivalent luminosity and redshift range (from Franceschini et al. 2003), we derived an age for those distant LIRGs of $t_{\rm form}=$ 5.5$\pm$1.5 Gyr, suggesting that they formed at $z_{\rm form}=$ 4.5$\pm$1.5.

As for the bursts themselves, we computed a median duration of $\tau_{\rm B}=1.0_{-0.6}^{+1.6}\times10^8$ years, during which the galaxies produced 10$\pm$9\,\% (the error bar includes 68\,\% of the galaxy sample) of their stellar mass. This corresponds to a mass of molecular gas of about 7$\times$10$^9$ M$_{\odot}$ which is consistent that observed in local LIRGs and ULIRGs (see Sanders \& Mirabel 1996). 

We note that all simulations produced in this paper assume the same fixed
IMF for both the underlying star formation and the burst of star
formation. Some evidence that the formation of low-mass may be less efficient in
the environment of active star formation in the solar neighborhood
were suggested in the past (Larson 1986, Scalo 1986, Maeder 1993). 
A top-heavy IMF could also
account for the enhanced ratio of light elements to iron in massive
early-type galaxies (Worthey, Faber \& Gonzalez 1992) and for the relative
enrichment of oxygen to iron in the intra-cluster medium (Arnaud et al,
1992). In our study, the occurrence
of a top-heavy IMF in the burst episode would only weakly influence the
derived burst duration timescales, which are set by the spectral signature
of massive A to F stars. However, a top-heavy IMF during the burst phase
would imply a much lower contribution to the total galaxy mass by
longer-lived, low-mass stars.

Finally, we discussed the possibility that the distant LIRGs experienced a previous starburst prior to the ongoing one during the last 1.5 Gyr. While most galaxies are not consistent with a merger scenario where two galaxies merge in several phases producing a series of bursts separated by a few ten millions years (Mihos \& Hernquist 1996), the majority present more than 50\,\% chance to have experienced a previous burst in the last 1.5 Gyr, i.e. since $z\sim$ 1. If these properties are typical of LIRGs between $z=$ 1 and $z=$ 0, then this suggests that the population of galaxies experiencing LIRG phases experienced on average 2 to 3 LIRG phases since $z=$ 1 and up to 4 since their birth around $z=$ 4--5, as also suggested by Hammer et al. (2005). 
This scenario is not consistent with the formation of distant LIRGs
through the continuous star formation characterizing isolated spiral
galaxies as has been argued independently based on their
morphology. Instead, minor mergers, tidal interactions and gas
accretion all remain equally plausible triggering mechanisms for more
than half of the distant LIRGs which do not harbor the morphology of
major mergers.

\acknowledgements{We wish to thank the anonymous referee for his constructive remarks which helped improving the paper in particular on the wavelet analysis. We also wish to thank Anna Gallazzi, Nicolas Gruel, Emmanuel Moy and Jean Luc Starck for helpfull discussions and comments and Emeric Le Floc'h for technical support for MIPS data. S.C. thanks the Alexander von Humboldt
Foundation, the Federal Ministry of Education and Research, and the
Programme for Investment in the Future (ZIP) of the German Government for
their support.}

\nocite{*}
\bibliography{biblio}

\end{document}